\documentclass[prd,aps,showpacs]{revtex4} 

\usepackage{amsmath}
\usepackage{graphicx}
 
\begin{document}

\title[Short Title]{Space--Time Symmetry, $CPT$ and Mirror Fermions}
\author{S. Ying} 
\email{sqying@yahoo.com,sqying@fudan.edu.cn}
\affiliation{Research Center for Theoretical Physics, Physics Department, 
         Fudan University,
          Shanghai 200433, China.}

\begin{abstract}
  The motivations for the construction of an 8-component
  representation of fermion fields based on a two dimensional
  representation of time reversal transformation and $CPT$ invariance
  are discussed. Some of the elementary properties of the quantum
  field theory in the 8-component representation are studied. It
  includes the space-time and charge conjugation symmetries, the
  implementation of a reality condition, the construction of
  interaction theories, the field theoretical imaginary- and real-time
  approach to thermodynamics of fermionic systems, the quantization of
  fermion fields, their particle content and the Feynman rules for
  perturbation theories. It is shown that in the new presentation, a
  $CPT$ violation can be formulated in principle. The construction of
  interaction theories in the 8-component theory for fermions is shown
  to be constrained by the $CPT$ invariance.  The short distance
  behavior and relativistic covariance are studied. In the path
  integral representation of the thermodynamical potential, the
  conventional imaginary-time approach is shown to be smoothly
  connected to a real-time thermal field theory in the 8-component
  representation for fermion fields without any additional subtraction
  of infinities.  The metastability at zero density and the nature of
  the spontaneous $CP$ violation in color superconducting phases of
  strong interaction ground states are clarified.
\end{abstract}

\maketitle

\section{Introduction}

 The role played by the space--time and charge conjugation symmetries
in fundamental interactions remain to be one set of the central
questions in modern physics. Among them, the time reversal
transformation (T) is a particularly subtle problem (see, e.g., Ref.
\cite{thesis}).  The contemporary edifice of quantum field theory
literally exclude the possibility of a systematic definition of the so
called causal reversal of time \cite{thesis}, which includes an
interchange of the initial and final states in a reaction. Only the so
called motion reversal of time \cite{thesis} that reverses only the
quantum numbers like the momentum, angular momentum, magnetic field,
etc. together with a complex conjugation can be defined. This is
because the particles are not allowed to travel backward in time in
quantum field theory, which specifies that the particles and their
antiparticles (and/or holes) have to travel forward in time. Other
reasons for the needs of such a revisit of the question of time
reversal are given in the main text. The above mentioned restriction
of quantum field theories maybe is of no physical significance if the
time reversal invariance is respected. But it is known that the time
reversal invariance should be violated in fundamental interactions
according to the observed $CP$ violation in the neutral kaon system
when $CPT$ is assumed to be an exact symmetry. The recent experimental
discoveries of the causal \cite{CPLEAR} and the motion \cite{KTeV}
time reversal invariance violations in the neutral kaon system make a
more detailed analysis of the problems involved in the time reversal
and the $CPT$ invariance timely.

The phenomenological implications of the theoretical suggestion that
$CP$ is spontaneously violated in the metastable (or virtual) color
superconducting phase of the strong interaction vacuum state
\cite{front,Paps1,ThPap,ThPap2} (see section \ref{app:ColorSuperc})
needs to be clarified concerning whether the time reversal or the
$CPT$ symmetry is kept in such a case. Therefore such an analysis is
carried out in the context of an understanding of relativistic matters
in particle/nuclear physics with finite matter and/or energy density
that underlies a variety of the currently interested topics in high
energy physics and astrophysics/cosmology.  A theoretical attempt is
proposed in Refs. \cite{ThPap,ThPap2,TMU}, which is adopted to
investigate the nucleon structure \cite{Nstru1,Nstru2,Nstru3}, the
nucleon stability and $\Sigma_N$ term problems \cite{Nstab} where
several of the novel ingredients of the local theory are shown to be
essential for a solution of the problems.  One of the basis of the
framework, namely the 8-component representation for fermions, has not
been discussed in enough details. Its relation to the enforcement of
$CPT$ invariance is not yet explicitly revealed. This work formalizes
the 8-component theory into a more complete quantum field theory.

A solution of the above mentioned problems related to time reversal is
expected to be found in a two dimensional representation of time
reversal transformation \cite{CPTt}. It should be put forward in a
more systematic and quantitative way in addition to the existing ones
in our earlier publications \cite{ThPap,ThPap2,Paps2,CPTt}. The
possibility of considering a two dimensional representation of time
reversal was discussed early by Wigner \cite{Wigner1}. Later works in
that direction were published in Refs. \cite{Weinbg,Chamb,Erdem}.  A
different realization of the two dimensional representation of the
time reversal transformation is derived in this work based on the
physical and conceptual requirements considered here.

There is another thread of reasoning that leads us to
the consideration of the 8-component representation for fermions.
Relative simple calculations reveal that new divergences in a quantum
field theory occur when it is used to handle finite density
problems. These additional divergences make the corresponding theory
ambiguous in such situations. It also violates the Lorentz
covariance. Arbitrary subtractions of those unwanted infinite
quantities need to be introduced to define a finite theory that is
physically sensible. But such a procedure reduces the predictive power
and spoils the Lorentz covariance of the theory. The question is can
such a problem be solved?

To deal with these type of new divergences, one can either stick to
the theoretical frame-work valid at zero density and straightforwardly
extend the theory to the finite density case by subtracting those
infinities regarded as non-physical. Such an approach has a long
history and it works.  The perturbation theory for interacting finite
density quantum field theories were fully developed in the past using
the 4-component representation for the fermions in the Euclidean
space-time. Related papers are too numerous to list here.  Some of the
standard problems are studied systematically and to great details. One
of the representatives of them can be found in, e.g.,
Refs. \cite{Freedman}. There are problems left unanswered by
these works concerning the fermion determinant at finite density in
the Minkowski space-time version of the theory. These problems are
related to how to subtract the infinite contributions of the
unphysical Dirac sea. It is found that the Dirac sea contributions can
not be subtract away once for all at the level of vanishing background
fields or zero density in the 4-component representation of the
fermions. They manifest themselves when one tries to make a connection
between the Euclidean and Minkowski version of the same theory at
finite density.

Or, one can, provided it is possible at all, embed certain new
theoretical structures into the existing framework of quantum field
theory in which some of old the divergences that one encounters in the
finite density cases are absent automatically.  The particular
candidate that is going to be studied here is the $CPT$ mirror partner
hypothesis \cite{CPTt} for each particle that solves the above
mentioned difficulties concerning time reversal transformation. The
theoretical gain is of at least two folds: the first
one is the reduction of independent hidden assumptions of the theory
and the second one is the possibility of making new physical
predictions that can be tested experimentally under conditions when
specific and/or concrete empirical knowledge is still lacking. Let us
for the moment value such an increase of the predictive power of the
theory more than a straightforward extension of theory that is only
well test at zero density and in non-relativistic situations by trying
to find out what an alternative, which is identical to the well
established ones at zero density and in non-relativistic situations,
can be derived and what is its implications.

 It is also quite interesting to see whether or not the resulting
quantum field theories derived following the above mentioned two
threads of reasoning are actually the same one.

 A two dimensional representation of the time reversal transformation
is introduced in section \ref{sec:8CompTheory}, which implies an
8-component representation for the fermion fields. The representation
of parity $P$, charge conjugation $C$ and $CPT$ transformations in the
8-component theory for fermions is discussed. The reality condition,
which is consistent with the $CPT$ invariance is discussed in more
details. The representation of any internal symmetries for the
fermionic particle in the 8-component theory is provided. The proper
fermion propagator that is consistent with the reality condition is
derived. The question of how to construct interaction terms in the
8-component theory for fermions is studied in the last part of this
section.  It is shown that $CPT$ violating theory can in principle be
constructed in the 8-component theory. In section \ref{sec:FDQFT},
another thread of reasoning that is in favor of the 8-component theory
is provided by doing a technical analysis of the problems of the
4-component theory for fermions as compared to the 8-component one in
the finite density situations when one tries to connect the Euclidean
form of the theory to its Minkowski correspondence. It is shown that
the divergent contributions contained in the 4-component theory is
absent in the 8-component one. The problems of the 4-component theory
is further exposed as one tries to relate local observables computed
in different Lorentz frame. It is shown that such a problem is absent
in the 8-component theory. In addition, the practically useful
framework of the imaginary- and real- time thermal field theory in the
8-component theory is established.  The quantization of the
8-component theory is provided in section \ref{sec:Particle} following
the formal rules of quantum field theory. The time reversal
transformation of the quantized 8-component fermion field operator is
discussed. The constraints of the reality condition on the creation
and annihilation operators for fermionic particles and antiparticles
is deduced.  An additional rule for elementary processes and
perturbation theories is derived from the reality condition.  In
section \ref{app:ColorSuperc}, the newly developed 8-component theory
for fermion is used to discuss two remaining issues concerning the
possible metastable color superconducting phase of the strong
interaction vacuum state. The metastability at zero density and the
CPT invariance of the color superconducting phase are established
based on the 8-component theory.  A summary and discussion is given in
section \ref{sec:Sum}.

\section{The 8-component Theory for Fermions}
\label{sec:8CompTheory}
\subsection{Space-time, charge symmetries and $CPT$ theorem}

   The reversal of time in quantum mechanics involves not only an
unitary transformation but also a complex conjugation \cite{Wigner2}.
This is because unitary representation of the time reversal
transformation is inconsistent with well established notions in
physics which require that the (space-time) coordinate and the
4-momentum of the system transform in an opposite way.  Such an
anti-unitary transformation is different from the ordinary unitary
transformations connected to other symmetries of the system in some
subtle ways, which has to be studied in more details to explore new
possibilities.

 For a particle with internal degrees of freedom corresponding to
certain ordinary symmetry transformations, the time reversal
transformation of it has to be so constructed as not to be
inconsistent with these symmetry transformations since the action of
complex conjugation transforms a particle belonging to a
representation ${\cal R}$ of certain symmetry group into the conjugate
representation $\overline {\cal R}$ of it, which is not always
equivalent to the original one. This causes conceptual problems since,
on the one hand, a particle belongs to a representation of a symmetry
group and its antiparticle belongs to the conjugate representation, on
the other hand, the time reversal transformation is designed to
generate from a particle state another state which is also a particle.
Thus the time reversed particle (or antiparticle) appears to has no
state to stay in!

Albeit on the practical level studied so
far, this conceptual problem rarely cause problems in observables that
are even under time reversal.  This is because the physical
observables like the currents always belong to the self-conjugate
adjoint representation of a symmetry group, which gives an equivalent
representation between the original and time reversed states.

It is still a challenge to search for a consistent way to solve
this conceptual problem related to our abstract notion of ``state''
for a particle (or antiparticle) by introducing a $CPT$ mirror state
for each relativistic particles that maps into one another under
the time reversal transformation. 

\subsubsection{Representation of the time reversal transformation}

The Lorentz group is covered by $SL_L(2,C)\times SL_R(2,C)$ with
$SL_{L/R}(2,C)$ the group of all complex $2\times 2$ matrices with
unit determinant. The spinor representation of the Lorentz group is
also the representation of the $SL_L(2,C)\times SL_R(2,C)$ group,
which, like the $SU(2)$ group, is self-conjugate with a common metric
tensor $g_{\alpha\beta}=(i\sigma_2)_{\alpha\beta}$ for both
$SL_L(2,C)$ and $SL_R(2,C)$. The the spinor representation of the
Lorentz group is provided by a pair of two component spinors, $\psi_L$
and $\psi_R$ that transform under the action of the matrices belonging
to $SL_L(2,C)$ and $SL_R(2,C)$ respectively.

  As far as the time reversal transformation is concerned, the Lorentz
spinors $\psi_L$ and $\psi_R$ behave in a similar way as the
2-component spinor for spatial rotation \cite{Landau}. If only one
pair of the Lorentz spinors $\psi_L$ and $\psi_R$ is considered, the
time reversal transformation should be
\begin{subequations}
\label{eq:T-rev-phiLR}
\begin{eqnarray}
         \phantom{{\bf x}}
         \psi_L^\alpha ({\bf x},t) 
                  &\stackrel{\widehat{\cal T}}{\longrightarrow}&
                \widetilde \psi_{L}^\alpha( {\bf x},t) =
e^{i\delta}g^{\alpha\beta}
         \psi^*_{L,\beta}({\bf x},-t),
\label{T-rev-phiL}\\
         \psi_{R}^\alpha({\bf x},t) &\stackrel{\widehat{\cal T}}{\longrightarrow}&
                \widetilde \psi_{R}^\alpha({\bf x},t) =
-e^{i\delta}g^{\alpha\beta}
         \psi^*_{R,\beta}({\bf x},-t),
\label{T-rev-phiR}
\end{eqnarray}
\end{subequations}
where $\delta$ is an arbitrary phase, in order to be consistent with
the Lorentz covariance. Here the superscript ``*'' denotes complex
conjugation.  A success of two time reversal transformations of
$\psi_{L/R}$ returns the negative of it no-matter what value $\delta$
takes.  In addition, one has the above mentioned problems associated
with the belonging of the representation for a time reversed particle
when the representation is not self-conjugate.

The solution to this problem, as it is discussed above, is to
introduce two spinors $\psi_1$ and $\psi_2$ to represent the time
reversal transformation in a way that is Lorentz covariant
\begin{subequations}
\label{eq:T-rev2-psiLR}
\begin{eqnarray}
         \psi_{1L}({\bf x},t) &\stackrel{\widehat{\cal T}}{\longrightarrow}&
                \widetilde \psi_{1L}({\bf x},t) = -
         e^{i\delta}
         \psi_{2R}({\bf x},-t),
                                           \label{T-rev-psi1L}\\
         \psi_{1R}({\bf x},t) &\stackrel{\widehat{\cal T}}{\longrightarrow}&
                \widetilde \psi_{1R}({\bf x},t) =
          e^{i\delta}
         \psi_{2L}({\bf x},-t),
                                           \label{T-rev-psi1R}\\
         \psi_{2L}({\bf x},t) &\stackrel{\widehat{\cal T}}{\longrightarrow}&
                \widetilde \psi_{2L}({\bf x},t) = 
         e^{i\delta}
         \psi_{1R}({\bf x},-t),
                                           \label{T-rev-psi2L}\\
         \psi_{2R}({\bf x},t) &\stackrel{\widehat{\cal T}}{\longrightarrow}&
                \widetilde \psi_{2R}({\bf x},t) =-
          e^{i\delta}
         \psi_{1L}({\bf x},-t).
\label{T-rev-psi2R}
\end{eqnarray}
\end{subequations}
The ``left'' and ``right''
handed $\psi_1$ and $\psi_2$ belong to mutually conjugate
representations of its internal symmetry groups, if present. It can be
seen that the complex conjugation involved in the one dimensional
representation of the time reversal transformation given by
Eqs. \ref{eq:T-rev-phiLR} is no longer explicit in the two component
representation defined in Eqs.  \ref{eq:T-rev2-psiLR}. After
performing a sequence of two transformations defined in
Eqs. \ref{eq:T-rev2-psiLR}, the spinor returns to its self multiplied
by a phase factor
\begin{equation}
  \chi({\bf x},t) \stackrel{\widehat {\cal T}\widehat {\cal T}} \longrightarrow  
                                     e^{2i\delta}\chi({\bf t},t)
\end{equation}
with $\chi$ any of the spinor in Eqs. \ref{eq:T-rev2-psiLR}.  Each
$\delta$ labels a different representation of the time reversal
transformation. The physically natural choice at the present is given
by $\delta=0$. But one should be open to the possible $\delta\ne 0$
physics, if any.

For the description of massive charged particles, Dirac spinor $\psi$,
which is 4 dimensional since it is formed via a combination of
$\psi_L$ and $\psi_R$, is used \cite{Ramond}. The metric tensor for
the Dirac spinor is the charge conjugation matrix $C$ with the
property
\begin{eqnarray}
       C^{-1} &=& C^\dagger = C^T = -C.
\end{eqnarray}
The Dirac spinor for a free massive particle satisfies the Dirac equation
\begin{eqnarray}
       \left ( i\rlap\slash\partial - m\right )\psi &=& 0,
\label{Dirac-eq}
\end{eqnarray}
where $m$ is the mass of the particle and
$\rlap\slash\partial=\gamma_\mu\partial^\mu$ with $\gamma^\mu$
one of the Dirac matrices.

The free Dirac equation respect the time reversal transformation
symmetry so that if $\psi({\bf x},t)$ is a solution of it, so is its
time reversed one defined by
\begin{eqnarray}
    \widetilde \psi({\bf x},t) = \gamma^5 C \psi^*({\bf x},-t),
\label{T-rev-psi}
\end{eqnarray}
which can be derived from Eq. \ref{eq:T-rev-phiLR} after a special choice
of the phase $\delta$ by putting $\delta=0$.

When the two Dirac spinors are used to represent the time reversal
transformation following the above reasoning, which is compactly
written as an 8 component spinor 
\begin{equation}
\Psi=\left (\begin{array}{c} \psi_1 \\ \psi_2 \end{array}
     \right ),
\end{equation}
where $\psi_{1,2}$ are 4-component spinor fields.  $\Psi$ transforms
in the same way as 4-component Dirac spinor $\psi_1$ or $\psi_2$
under the Poincar\'{e} group. It satisfies the same equation as the
Dirac spinor, namely,
\begin{eqnarray}
     \left ( i\rlap\slash\partial - m\right )\Psi &=& 0
\label{Dirac-eq-Psi}
\end{eqnarray}
with $\psi_1$ and $\psi_2$ conjugate to each other in the non
self-conjugate representations of internal symmetry groups.

Then the time reversal transformation different from the one given in
Eq. \ref{T-rev-psi} is derived from the elementary ones in
Eq. \ref{eq:T-rev2-psiLR}
\begin{eqnarray}
      \Psi({\bf x},t) &\stackrel{\widehat{\cal T}}{\longrightarrow}&
             \widetilde \Psi({\bf x},t) = \gamma^0\gamma^5 iO_2
\Psi({\bf x},-t),
\label{T-rev-8}
\end{eqnarray}
where $\delta=0$ is assumed and $O_2$ is the second of the three Pauli
matrices $O_{1,2,3}$ acting on the upper ($\psi_1$) and lower
($\psi_2$) 4 components of $\Psi$. It can be shown that a sequence of
two time reversal transformations on $\Psi({\bf x},t)$ returns to
$\Psi({\bf x},t)$ instead of negative of it.

The transformation of fermion bilinear operators can be deduced after
the transformation of $\Psi$ is known. For example, for c-number currents
of the path integration formulation
\begin{subequations}
\label{eq:vec-T-trans}
\begin{eqnarray}
V_I^\mu (x)  &=&\overline\Psi({\bf x},t)\gamma^\mu O_3\Psi({\bf x},t) 
      \stackrel{\widehat{\cal T}} {\longrightarrow} -
       \overline\Psi({\bf x},-t)\gamma_\mu O_3 \Psi({\bf x},-t),
\label{vec-p-tt} \\
V_{II}^\mu (x) &=&\overline\Psi({\bf x},t)\gamma^\mu \Psi({\bf x},t) 
      \stackrel{\widehat{\cal T}}{\longrightarrow} 
\overline\Psi({\bf x},-t)\gamma_\mu \Psi({\bf x},-t).
\label{vec-m-tt} 
\end{eqnarray}
\end{subequations}

 It must be emphasized that although there is no explicit complex conjugation
in Eq. \ref{T-rev-8}, the time reversal transformation is still an anti-linear
transformation. Any ordinary number that does not carry the Dirac index is 
complex conjugated or transposed (see the following), namely if a 8-component
spinor $\Psi$ can be expressed as a linear combination of several other ones
\begin{equation}
 \Psi = \sum_n c_n \Phi_n
\end{equation}
then 
\begin{equation}
\Psi \stackrel{\widehat{\cal T}}{\longrightarrow} \widetilde \Psi =
\sum_n c'_n \widetilde \Phi_n,
\end{equation}
where
\begin{equation}
\Phi \stackrel{\widehat{\cal T}}{\longrightarrow} \widetilde \Phi. 
\end{equation}
The coefficient $c'_n= c^*_n$ if it is a c--number. It will be shown
in the following that if $c_n$ is an operator of the quantum theory,
there are two ways to represent the anti-linear transformation: either
$c'_n=c^*_n$ or $c'_n=c^T_n$ with the superscript ``T'' denoting the operator
transpose. These two representations are not always equivalent.

\subsubsection{The parity $P$ and charge conjugation $C$ transformations}

   The parity transformation $\widehat {\cal P}$ is performed on the upper and
lower 4 components of $\Psi({\bf x},t)$ separately
\begin{eqnarray}
     \Psi({\bf x},t)  &\stackrel{\widehat{\cal P}}\longrightarrow & \gamma^0 O_3 
      \Psi(-{\bf x},t).
\label{Psi-P-tra}
\end{eqnarray}

   The charge conjugation transformation $\widehat{\cal C}$ is also
performed on the upper and lower 4 components of $ \Psi({\bf x},t)$
separately
\begin{eqnarray}
      \Psi({\bf x},t) &\stackrel{\widehat{\cal C}}\longrightarrow& 
                 C \gamma^0 \Psi^*({\bf x},t).
\label{Psi-C-tra}
\end{eqnarray}
\subsubsection{The $CPT$ transformation} 
 
The anti-linear $\widehat{\cal CPT}$ transformation of the
8-component fermion field is then given by
\begin{equation}
     \Psi({\bf x},t) \stackrel{\widehat{\cal CPT}}
      \longrightarrow  \gamma^5 C O_1 \gamma^0
     \Psi^*(-{\bf x},-t).
\label{Psi-CPT-tra}
\end{equation}

  There are more ways to break the fundamental
$CP$, $T$ and even $CPT$ symmetry in the 8-component representation
for the fermion field.  For example, it is shown \cite{ThPap,ThPap2}
that the time component of the statistical gauge field $\mu^\alpha$
has a non-vanishing vacuum expectation value in the possible
metastable color superconducting phases of the strong interaction
vacuum state. Since $\mu^\alpha$ is coupled to $V_{I}^\alpha$ defined
in Eq. \ref{vec-m-tt} in the local theory \cite{ThPap,ThPap2}, such a
non-vanishing vacuum $\mu^0$ violates both $CP$ and $T$ invariance while
keeps the $CPT$ invariance.  This is discussed in section
\ref{app:ColorSuperc}. If the statistical gauge field is coupled to
$V_{II}^\alpha$ instead, then a non-vanishing vacuum $\mu^0$ violates
$CP$ and $CPT$ invariance but keeps $T$ invariance. The reasons why
the statistical gauge field (the chemical potential for a uniform
system) should be coupled to $V_{I}^\mu$ is given in the following.

\subsection{Other internal symmetries}

It is discussed above that the upper 4-component and lower 4-component
of $\Psi$ belong to mutually conjugate representation of internal
symmetry groups.  Suppose the generators for a group ${\cal G}$ that
has non self-conjugate representations are $t_a$ ($a=1,2,\ldots,n$),
which satisfy the following commutation relation
\begin{eqnarray}
      \left [t_a,t_b\right ] &=& f^{abc} t^c
\label{G-comm-rel}
\end{eqnarray}
with $f^{abc}$ the structure constants of ${\cal G}$.  Let the spinors
 $\psi_1$ belongs to a representation ${\cal R}$ of it and $\psi_2$
 belongs to a conjugate representation $\bar{\cal R}$ of ${\cal
 R}$. The infinitesimal transformation of $\psi_1$ and $\psi_2$ under
 ${\cal G}$ is
\begin{subequations}
\label{eq:G-trans}
\begin{eqnarray}
    \delta \psi_1^m &=& i \delta\omega^a t^m_{a,n}
    \psi_1^n,\label{G-trans-1}\\
    \delta \psi_{2,m} &=& - i \delta\omega^a t^n_{a,m}
    \psi_{2,n}\label{G-trans-2},
\end{eqnarray}
\end{subequations}
where $\delta\omega^a$ are infinitesimal group parameters. Then the spinor
$\Psi$ introduced above transforms in the following way
\begin{eqnarray}
   \delta \Psi^m &=& i\delta\omega^a T^m_{a,n} \Psi^n
\label{G-trans-chi}
\end{eqnarray}
with generator $T_a$ defined by
\begin{eqnarray}
    T_a &=& \frac 1 2 (1+O_3) t_a - \frac 1 2 (1-O_3) t^T_a,
\label{T-def}
\end{eqnarray}
where superscript ``T'' denotes transpose. Since $t_a$ is either
symmetric or antisymmetric for an unitary representation, it can be
shown that $T_a$ satisfy the same commutation relation as the original
$t_a$ given by Eq. \ref{G-comm-rel}, which provides an adjoint
representation of ${\cal G}$.

\subsection{The reality condition}

 Since the 8-component field contains twice excitation modes compared
to the physical modes contained in the usual 4-component
representation, there are redundant modes in the 8-component
one. Certain forms of a pairwise identification of the excitation
modes of the 8-component theory has to be searched for at zero density
in which case the 4-component representation for fermion fields is
thoroughly tested in observations. The straight forward identification
of the positive (negative) energy solutions to the Dirac equation
Eq. \ref{Dirac-eq-Psi} can be shown to lead to inconsistencies. 

It is found that the proper identification of equivalent excitation
modes at zero density can be implemented at the field level based on
the fundamental $CPT$ invariance, which is called the reality condition
\cite{ThPap,ThPap2}. 

\subsubsection{Two kinds of mirror reflection operation}

Let us introduce two anti-linear mirror reflection operations
$\widehat {\cal M}$ and $ \widehat {\cal R}_{\cal M}$ which are used
to express the reality conditions for fermion fields as well as the
boson fields.  Their realization on the fermion fields are
\begin{equation}
  \Psi(x) \stackrel{\widehat {\cal M}}\longrightarrow {\cal M} 
         \overline\Psi^T(-x)  
\label{M-def}
\end{equation}
with superscript ``T'' denoting transpose and
\begin{equation}
  \Psi(x) \stackrel{\widehat {\cal R}_{\cal M}}\longrightarrow {\cal R}_{\cal M} 
         \Psi^*(-x).  
\label{RM-def}
\end{equation}
The representation matrix ${\cal M}$ is
\begin{equation}
      {\cal M} = O_1 C.
\end{equation}
It follows that the representation matrix ${\cal R}_{\cal M}$ is  
\begin{equation}
   {\cal R}_{\cal M} = {\cal M}\gamma^0 = \gamma^5 CPT.
\label{M-CPT}
\end{equation}

Note both $\widehat {\cal R}_{\cal M}$ and $\widehat {\cal M}$ contain time
reversal transformation, they are anti-linear operators. In the
quantized theory $\widehat{\cal R}_{\cal M}$ is derived from the so called
motion reversal of time transformation; it involves a complex
conjugation.  $\widehat{\cal M}$ is derived from the so called causal
reversal of time; its effects on Grassmanian numbers or creation and
annihilation operators is represented by a transpose instead of
complex conjugation. This will be discussed in subsection
\ref{sec:qTRVT}. Therefore they are different operations in
principle. Therefore Eq. \ref{M-CPT} should be understood as
considering only the unitary part of the time reversal transformation.

 The transformation of the boson fields under these two kinds of
mirror reflection operation is constrained by the $CPT$ invariance and
the kind of dynamics involved. Since boson field can be represented by
bilinear form in terms of anticommuting fermion field, these two mirror
reflection differs by a sign for real classical field (see subsection
\ref{sec:bMirr}).

\subsubsection{The reality condition for fermions}

The reality condition is given by \cite{Paps2,ThPap}
\begin{eqnarray}
  \Psi(x) = {\cal M} \overline\Psi^T(-x)= \gamma^5
    \Psi_{CPT}(x),
\label{realcond}
\end{eqnarray}
where $\Psi_{CPT}(x)$ is the image of $\Psi(x)$ under the $CPT$ transformation.

\subsection{The fermion propagator}

\subsubsection{Zero density case}

 In case of a free fermionic system, its Lagrangian density is a
bilinear form of the fermion field. A tentative form of it would be,
\begin{equation}
     {\cal L}_0= \frac 1 2 \overline \Psi 
             \left ( i\rlap\slash\partial - m \right ) \Psi.
\label{fact}
\end{equation}
The momentum space Dirac equation derived from it has four sets of
solutions.  The first one is the positive energy solution with all
lower 4 components vanish, which is denoted as
$U^{(+)}_1(\mathbf{p})$. The second one is the negative energy
solution with all lower 4 components vanish, which is denoted as
$U^{(-)}_1(\mathbf{p})$. The third one is the positive energy solution
with all upper 4 components vanish, which is denoted as
$U^{(+)}_2(\mathbf{p})$.  The fourth one is the negative energy
solution with all upper 4 components vanish, which is denoted as
$U^{(-)}_2(\mathbf{p})$.  They are not all independent. So one can not
simply invert the operator inside the bracket of Eq. \ref{fact} to get
the singular fermion propagator. Proper arrangement of these c--number
solutions has to be made so that the equivalent solutions that can be
mapped into each other by the mirror reflection operation
\begin{eqnarray}
  \Psi(x) \longrightarrow {\cal R}_{\cal M}\Psi^*(-x).
\label{realtrans}
\end{eqnarray}

The reality condition requires that the momentum space eigenvalue
problem
\begin{equation}
      \gamma^0 \left ( \rlap\slash p-\Sigma \right ) \Psi(\mathbf{p}) 
         = \lambda \Psi(\mathbf{p}),
\label{EigEq}
\end{equation}
which is relevant to the path integral calculation of the effective
action for fermions \cite{ThPap,ThPap2}, to be invariant under the
mirror transformation Eq. \ref{realtrans}, namely
\begin{equation}
       {\cal R}_{\cal M}\gamma^0 \left ( \rlap\slash p-\Sigma \right )^* 
        {\cal R}^{-1}_{\cal M}
        = \gamma^0 \left ( \rlap\slash p-\Sigma \right ). 
\label{realcond1}
\end{equation}

It leads to a constraint for the form of the ``mass'' term, namely
$\Sigma$ must satisfy the following equation
\begin{equation}
       {\cal R}_{\cal M} \gamma^0 \Sigma^* {\cal R}^{-1}_{\cal M}
        = \gamma^0 \Sigma. 
\label{realcond2}
\end{equation}
In the simplest case of free theory, one must write
\begin{equation}
 \Sigma = \left ( \begin{array}{cc} m & 0 \\ 0 & -m \end{array}
          \right ) = m O_3.
\label{nsc-mass}
\end{equation} 
It tells us that one has to make the following identification 
\begin{eqnarray} 
U^{(+)}_1(\mathbf{p})&\sim & U^{(-)}_2(\mathbf{p}), \\
U^{(-)}_1(\mathbf{p})&\sim & U^{(+)}_2(\mathbf{p})
\end{eqnarray} 
for the solution of Dirac Eq. \ref{fact}
instead of other ones. The upper 4 components and the lower 4 components
of $\Psi$ are to be used to describe the same physics at zero chemical potential
since the mirror reflection ${\cal R}_{\cal M}$ maps them into each other.

The fermion propagator for a free fermion in the momentum space at
zero chemical potential and temperature, which is related to the
inverse of the operator on the left hand side of Eq. \ref{EigEq}
without $\gamma^0$, is
\begin{equation}
   S_F(p) = \frac {i}{\rlap\slash p - m O_3 + i\epsilon }.
\label{fprop}
\end{equation}
It is different from a formal inversion of the singular operator
$i\rlap\slash\partial - m$ in the momentum space. Therefore Eq.
\ref{fact} is more appropriately be written in the following equivalent 
from
\begin{equation}
     {\cal L}_0= \frac 1 2 \overline \Psi 
             \left ( i\rlap\slash\partial - m O_3 \right ) \Psi.
\label{fact1}
\end{equation}
Since the upper and lower 4 components do not couple to each other
and the sign of the mass term do not has physical significance in
relativistic theories, the single particle solutions of Lagrangian
density Eqs. \ref{fact} and \ref{fact1} are in one to one
correspondence. Therefore the
Lagrangian density can be expressed as Eq. \ref{fact} before the
reality condition Eq. \ref{realcond} is enforced. The reality
condition puts further constraints so that the propagator for the
fermions in the quantized theory is given by Eq. \ref{fprop}.

\subsubsection{The propagator in finite density situations}

The Lagrangian density for finite density situations in the local 8-component
theory \cite{ThPap,ThPap2} is
\begin{equation}
     {\cal L}_0= \frac 1 2 \overline \Psi 
             \left ( i\rlap\slash\partial + 
          \rlap\slash \mu  O_3 - m O_3 \right ) \Psi,
\label{local-fd-fact}
\end{equation}
where $\mu^\alpha$ is the statistical gauge field. Instead of getting
into the details of the local theory, we shall consider simpler case
of the corresponding global theory for uniform systems in which
$\mu^0$ corresponds to the chemical potential $\mu$. The global
version of Eq. \ref{local-fd-fact} is
\begin{equation}
     {\cal L}_0= \frac 1 2 \overline \Psi \left ( i\rlap\slash\partial
             + \mu \gamma^0 O_3 - m O_3 \right ) \Psi.
\label{fd-fact}
\end{equation}
The corresponding eigenvalue equation for the effective action is
\begin{equation}
      \gamma^0 \left ( \rlap\slash p+\mu\gamma^0 O_3-m O_3 \right ) 
         \Psi(\mathbf{p})  = \lambda \Psi(\mathbf{p}).
\label{EigEq1}
\end{equation}
It is changed to 
\begin{equation}
      \gamma^0 \left ( \rlap\slash p-\mu\gamma^0 O_3-m O_3 \right ) 
         \Psi(\mathbf{p}) = \lambda \Psi(\mathbf{p})
\label{EigEq2}
\end{equation}
under the mirror reflection Eq. \ref{realtrans}.  Although the
individual eigenvalues $\lambda$ are changed under the mirror
reflection Eq. \ref{realtrans}, the effective action is not
changed. The reason is that the effective action, which is
proportional to the logarithmic of the product of all eigenvalues of
Eq. \ref{EigEq1}, is expected to be an even function of $\mu$.

A comparison of Eq. \ref{EigEq1} and Eq. \ref{EigEq2} also tells us
that the mirror symmetry at zero chemical potential is broken by the chemical
potential. It makes the 8-component theory inequivalent to the
conventional 4-component one.

The fermion propagator at finite density is then the inverse of 
$(\rlap\slash p+\mu\gamma^0 O_3 - m O_3)$
\begin{equation}
   S_F(p) = \frac {i}{\rlap\slash p+\mu\gamma^0 O_3 - m O_3 + i\epsilon }.
\label{fd-fprop}
\end{equation}

\subsection{The construction of interacting theories}

\subsubsection{Lorentz covariant vertices and their $CPT$ transformation}

Fermions interact with each other by exchanging intermediate
bosons in fundamental local theories. Due to the Lorentz invariance, the type of their interaction
can be classified according to the spin of the bosonic particles been
exchanged. The possible bosons that can be exchanged are scalars
$\sigma$, pseudo-scalars $\pi$, vectors $v^\mu$, axial-vectors $a^\mu$
and tensors $k^{\mu\nu}$ with various internal symmetry. The
corresponding fermion ``current'' are
\begin{equation}
   S^a_I = \frac 1 2 \overline \Psi T^a\Psi, \hspace{1cm} 
   S^a_{II} = \frac 1 2 \overline \Psi T^a O_3\Psi, 
\label{scalar-v}
\end{equation}
which are Lorentz scalars with $T^a$ the generator of any internal symmetry
defined in Eq. \ref{T-def},
\begin{equation}
   P^a_I = \frac 1 2    \overline \Psi i\gamma^5 T^a\Psi, \hspace{1cm} 
   P^a_{II} = \frac 1 2 \overline \Psi i\gamma^5 T^a O_3\Psi, 
\end{equation}
which are Lorentz pseudo-scalars,
\begin{equation}
   V^{a\mu}_I = \frac 1 2    \overline \Psi \gamma^\mu  T^a\Psi, \hspace{1cm} 
   V^{a\mu}_{II} = \frac 1 2 \overline \Psi \gamma^\mu T^a O_3\Psi, 
\end{equation}
which are Lorentz vectors,
\begin{equation}
   A^{a\mu}_I = \frac 1 2    \overline \Psi \gamma^\mu \gamma^5 T^a O_3\Psi, 
                     \hspace{1cm} 
   A^{a\mu}_{II} = \frac 1 2 \overline \Psi \gamma^\mu \gamma^5 T^a\Psi, 
\end{equation}
which are Lorentz axial-vectors and
\begin{equation}
   T^{a\mu\nu}_I = \frac 1 2    \overline \Psi \sigma^{\mu\nu} T^a O_3 
           \Psi, \hspace{1cm} 
   T^{a\mu\nu}_{II} = \frac 1 2 \overline \Psi \sigma^{\mu\nu} T^a\Psi, 
\end{equation}
which are Lorentz tensors. Here $T^a=O_3$ if the internal symmetry is $U(1)$.

They have different transformation properties under
$CPT$. It can be shown that according to Eq. \ref{Psi-CPT-tra} 
\begin{subequations}
\label{eq:CPT-tra}
\begin{eqnarray}
   S^a_I(x) \stackrel{\widehat{\cal CPT}}\longrightarrow S^a_I(-x), &\hspace{1cm}&
   S^a_{II}(x) \stackrel{\widehat{\cal CPT}}\longrightarrow - S^a_{II}(-x),\\
   P^a_I(x) \stackrel{\widehat{\cal CPT}}\longrightarrow P^a_I(-x), &\hspace{1cm}&
   P^a_{II}(x) \stackrel{\widehat{\cal CPT}}\longrightarrow - P^a_{II}(-x),\\
   V^{a\mu}_I(x) \stackrel{\widehat{\cal CPT}}\longrightarrow V^{a\mu}_{I}(-x), 
           &\hspace{1cm}&
   V^{a\mu}_{II}(x) \stackrel{\widehat{\cal CPT}}\longrightarrow - V^{a\mu}_{II}(-x),\\
   A^{a\mu}_I(x) \stackrel{\widehat{\cal CPT}}\longrightarrow A^{a\mu}_{I}(-x), 
           &\hspace{1cm}&
   A^{a\mu}_{II}(x) \stackrel{\widehat{\cal CPT}}\longrightarrow - A^{a\mu}_{II}(-x),\\
   T^{a\mu\nu}_I(x) \stackrel{\widehat{\cal CPT}}\longrightarrow T^{a\mu\nu}_I(-x), 
   &\hspace{1cm}&
   T^{a\mu\nu}_{II}(x) \stackrel{\widehat{\cal CPT}}\longrightarrow - T^{a\mu\nu}_{II}(-x).
\end{eqnarray}
\end{subequations}

Due to the existence of two types of fermion bilinear operators under
the $CPT$ transformation for each representation of the Lorentz group,
it is possible to write down Lorentz covariant interaction terms that
violates the $CPT$ invariance.  Therefore the $CPT$ invariance is not
a theorem but a physical law in the 8-component representation for
fermions. For example, similar to the fact that a mix of vector and
axial-vector current leads to a violation of parity invariance, a
current of the following form
\begin{equation}
     \widetilde V^{a\mu}= \alpha V_I^{a\mu} + \beta V_{II}^{a\mu} 
\end{equation}
with neither $\alpha=0$ nor $\beta=0$ that couples to a bosonic field
$v_{a\mu}$ would lead to a violation of $CPT$ symmetry. In fact any
mixing of ``type--I'' operator and ``type--II'' operator within the
same representation of the Lorentz group implies a violation of the
$CPT$ invariance. Such possibilities will be studied in other works.

Given the fact that the $CPT$ invariance violation is not observed so
far, one can choose either the ``type--I'' current or the ``type--II''
operators to build interaction theories but not both. Due to the wide
success of the 4-component theory for fermion in most of the
situations considered, a choice can be made by requiring that the
results of the 8-component theory reduce to the 4-component theory in,
e.g., non--relativistic limits. As it is shown in the following, the
``type--I'' operators listed above are the suitable ones.  The
``type--II'' operators listed above lead to significantly different
physics from the predictions of the 4-component theory and in
addition, they have divergent ground state expectation values in a
thermo environment that are absent if the ``type--I'' operators are
chosen. In some sense they are bad behaviored operators due to their
divergent ground state expectation values. But if they are going to be
kept, the mixing of the ``type--II'' operators to the dominant
``type--I'' operators, if any, should be renormalized to have
effects that lie below the current experimental upper bounds for the
$CPT$ violation.

Therefore the general $CPT$ preserving chiral symmetric 
interaction term without the tensor term is
\begin{equation}
{\cal L}_{int} = g_s \left (\sigma_a S^a_I + \pi_a P^a_I \right ) + 
             g_v v_{a\mu}
             V_I^{a\mu} + g_a a_{a\mu} A_I^{a\mu},
\label{intLag}
\end{equation}
where $\sigma_a$, $\pi_a$, $v_{a\mu}$ and $a_{a\mu}$ are corresponding
boson fields with $g_s$, $g_v$ and $g_a$ the coupling constants.

 In addition, the 8-component theory allows the construction of more
general interaction theory if the $\{1,O_3\}$ matrices in between the
two fermion fields are replaced by $\{O_1,O_2\}$ matrices. The same
classification can be done along the same line. Two special cases are
discussed in section \ref{app:ColorSuperc}. As it is demonstrated in
there that if the interaction vertices are contracted from
$\{O_1,O_2\}$, then ``type--II'' operators are allowed.

\subsubsection{The mirror transformation of bosons}
\label{sec:bMirr}

The representation of the anti-linear mirror transformations
$\widehat{\cal M}$ and $\widehat{\cal R}_{\cal M}$ (or time reversal
transformation) on the integer spin boson fields does not require a
doubling of the degrees of freedom. It can be shown that the
``type--I/type--II'' $CPT$ invariant interaction requires that the
reality condition for the corresponding boson field to be given by
\begin{eqnarray}
   \phi(x) &\stackrel{\widehat {\cal M}}\longrightarrow& \pm (-1)^{\nu+1}
                 \phi^T(-x), \label{RC-boson1}\\
   \phi(x) &\stackrel{\widehat {\cal R}_{\cal M}}
      \longrightarrow& \pm (-1)^{\nu} \phi^*(-x) \label{RC-boson2} 
\end{eqnarray}
respectively, where $\nu$ is the number of Lorentz indices that the
boson field $\phi$ carry and the superscript ``T'' denotes transpose
if $\phi$ is treated as an operator (the reason is given in subsection
\ref{sec:qrealcond}).

\subsubsection{Perturbation theory and Feynman rule}

The Feynman rules for fermion systems with interaction are given in Ref. 
\cite{Paps2}. Care should be taken that in the 8-component theory
the following four kinds of contractions should be taken into account, 
namely,
\begin{eqnarray}
  S_F(x,y)=<0| T\Psi(x)\overline \Psi(y)|0>, &\hspace{1cm} &
  S^a_F(x,y)=<0| T\Psi(x)\Psi^T(y) |0>, \\
  S^b_F(x,y)=<0| T\overline \Psi(x)\overline \Psi(y)|0>, &\hspace{1cm} &
  S^c_F(x,y)=<0| T\overline \Psi^T(x)\Psi^T(y) |0>.
\end{eqnarray}
They are related to each other via the reality condition Eq. \ref{realcond}
\begin{equation}
  S^a_F(x,y) = -S_F(x,-y){\cal M}, \hspace{0.4cm}
  S_F^b(x,y) = -{\cal M} S_F(-x,y), \hspace{0.4cm}
  S^c_F(x,y) = {\cal M} S_F(-x,-y) {\cal M}.
\end{equation}
All the interaction vertices satisfy \cite{Paps2} 
\begin{equation}
     {\cal M} V_M^T(x) {\cal M}^{-1} = V(-x).
\label{RC-vert}
\end{equation}
Note that the vertex $V(x)$ here contains the corresponding boson
fields and the subscript ``M'' denotes that the corresponding boson
field is transformed according to Eqs. \ref{RC-boson1} and
\ref{RC-boson2}. Using these relations, one can transform all
different contractions listed above into the one that contains $S_F$
only. Another rule for writing down Feynman diagrams in perturbation
theory is found after the quantization of the theory is studied. It is
discussed in subsection \ref{sec:FR-q}.

\section{The Ground State Properties and Thermodynamics}
\label{sec:FDQFT}

Thermodynamics is the theory used to study the bulk properties of the
system at equilibrium. In the zero temperature limit, it provides a
natural mean to tackle the properties of the lowest energy state of
the system, namely the ground state.  The description of equilibrium
systems based on quantum field theory is called thermal field theory
in the following.

\subsection{Imaginary-time formulations of the thermal field theory}

One form of the thermal field theory was developed in the Euclidean
space-time \cite{Matsu} with the time variable playing the role of the
inverse temperature. Such a formalism, which is based on the periodic
(antiperiodic) boundary condition for bosons (fermions) fields, is
well known. It is reviewed here based on the conventional 4-component
theory since it provides a reference for isolate the potential
problems to be discussed in the next section.

\begin{figure}[ht]
\includegraphics{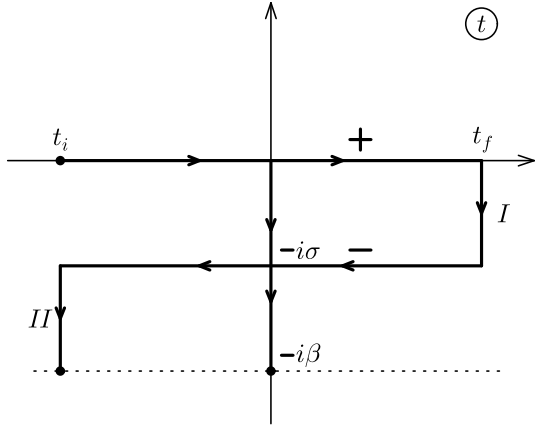}
\caption{\label{fig:RTcnt}
The contour for the time integration in the complex time
plane for equilibrium systems. The value of $\sigma$ is usually chosen
to be $\beta/2$. The initial time $t_i\to -\infty$ and the final time
$t_f\to\infty$.}
\end{figure}

The periodic (antiperiodic) boundary condition for the field operators
is explicitly written as
\begin{eqnarray}
 \widehat\psi(t-i\beta) & = &\xi
e^{-\beta\mu}\widehat\psi(t),\label{Fermion_BC}
\end{eqnarray}
where $\beta$ is the inverse temperature, $\xi=\pm 1$ for bosons and
fermions respectively and $\mu$ is the chemical potential.  The
physical responses of the system at equilibrium to small external
real-time stimulations can be evaluated after a
proper analytic continuation. Another formulation of the same
problem based upon Eq. \ref{Fermion_BC} can be derived by a distortion
of the Matsubara time contour, which goes straightly from $0$ to
$-i\beta$, to a contour that contains the real-time axis extending
from negative infinity to positive infinity and returning contour
below the real-time axis somewhere between $0$ and $-i\beta$ that
parallels the real-time axis (see Fig. \ref{fig:RTcnt}). Such a
real-time approach is discussed in the next subsection in more
details.

The thermodynamics of a system with variable fermion number density is
determined by the grand potential $\Omega$ defined as
\begin{eqnarray}
  e^{-\beta \Omega} &=& Tr e^{-\beta(\widehat H - \mu \widehat N)},
\label{PartFun1}
\end{eqnarray}
where $\widehat N$ is the fermion number and $\widehat H$ is the
Hamiltonian of the system. The grand potential $\Omega$ has a well
known relation with other thermodynamical quantities in elementary
thermodynamics, namely
\begin{equation}
\Omega = U-TS-\mu N 
\label{therm1}
\end{equation}
with $U$ the internal energy , $T$ the temperature, $S$ the entropy and
$N$ the average particle number of the system.

In the quantum field theoretical investigation of the same system,
there is another representation for the right hand side (r.h.s.)  of
Eq. \ref{PartFun1} in terms of Feynman--Matthews--Salam (FMS)
\cite{FMS} path integration
\begin{eqnarray}
  Tr e^{-\beta \widehat K} = const\times \int
  [D\psi][D\bar\psi] e^{-S_E},
\label{PartFun2}
\end{eqnarray}
where $\widehat K\equiv \widehat H - \mu \widehat N$, ``$const$'' is
so chosen that $\Omega=0$ at zero temperature and density. $S_E$ is
the Euclidean action of the system; it can be obtained from the
Minkowski action for the system by a continuous change of the metric
\cite{Mehta}. For example, the action $S_E$ for a free fermion system
is
\begin{eqnarray}
  S_E = \int d^4 x \bar\psi \left ( i\rlap\slash\partial + \mu
  \gamma^0_E - m \right ) \psi,
\label{EuclAct1}
\end{eqnarray}
where $x^\mu$ is the (4-dim) Euclidean space-time coordinates,
$\gamma_E^0=i\gamma^5$ and $m$ is the mass of the fermion. $\Omega$ is
obtained by identifying $\beta$ with the interval $x_4$ (the Euclidean
time) that the system is anti-periodic.

Since only ``free'' systems in which the Lagrangian for the fermion
system is a bilinear form in the fermion field are considered in this
section, the path integration can be carried out immediately.  The
standard procedure of doing the (imaginary) energy integration at zero
temperature is replaced by a summation of the discrete Matsubara
frequencies (energies) implied by Eq. \ref{Fermion_BC}. Such a
boundary condition for $\psi$ results, however, in different analytic
structure for the path integration from that for the FMS path
integration in the real-time in the 4-component theory for
fermions. This will be discussed in the following.

The grand potential in the Matsubara formalism is
\begin{eqnarray}
  \Omega &=& -\mbox{Sp}\mbox{Ln}\gamma_E^0 \left (i\rlap\slash\partial +
      \mu\gamma_E^0-m \right ) - \Delta\Omega \nonumber \\ &=&
   - \frac {2V}{\beta}
          \int \frac {d^3p} {(2\pi)^3} 
  \sum_{n=-\infty}^{\infty}\ln\left \{
  m^2+\mbox{\boldmath{$p$}}^2-\left [ i(2n+1)\pi\beta^{-1}+\mu \right
    ]^2 \right \} - \Delta\Omega,
\label{MatsuAct}
\end{eqnarray}
where ``Sp'' denotes the functional trace and $\Delta\Omega$ is so
chosen that $\Omega(T=\mu=0)=0$.  The summation over Matsubara frequencies
can be evaluated by expressing it as a contour integration in the
complex energy plane, namely,
\begin{equation}
  \Omega =- 2 V \int \frac {d^3p} {(2\pi)^3} \int_{C_0} \frac{dz}{2\pi i}
  \left [\mbox{tanh} (\frac 12 \beta z)  f(z,\mu)-\lim_{\beta\rightarrow \infty}
         \mbox{tanh}
          (\frac 12 \beta z) f(z,0) \right ],
\label{EuclActC0}
\end{equation}
where $C_0$ is shown in Fig. \ref{Fig:Matsu} and  
\begin{equation}
  f(z,\mu ) = \ln \left [ m^2+\mbox{\boldmath{$p$}}^2 - (z+\mu)^2 \right ].
\end{equation}
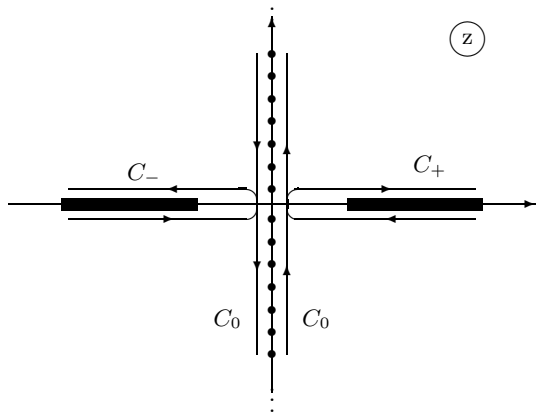
\begin{figure}[h]
 \unitlength=1.00mm \linethickness{0.5pt}
\begin{picture}(110.00,107.00)(20,10)
\put(45.00,80.00){\vector(1,0){70.00}}
\put(80.00,55.00){\vector(0,1){50.00}}
\put(80.00,60.00){\circle*{1.10}}
\put(80.00,63.00){\circle*{1.10}}
\put(80.00,66.00){\circle*{1.10}}
\put(80.00,69.00){\circle*{1.10}}
\put(80.00,72.00){\circle*{1.10}}
\put(80.00,75.00){\circle*{1.10}}
\put(80.00,78.00){\circle*{1.10}}
\put(80.00,82.00){\circle*{1.10}}
\put(80.00,85.00){\circle*{1.10}}
\put(80.00,88.00){\circle*{1.10}}
\put(80.00,91.00){\circle*{1.10}}
\put(80.00,94.00){\circle*{1.10}}
\put(80.00,97.00){\circle*{1.10}}
\put(80.00,100.00){\circle*{1.10}}
\put(90.00,79.20){\rule{18.00\unitlength}{1.60\unitlength}}
\put(52.00,79.20){\rule{18.00\unitlength}{1.60\unitlength}}
\put(82.00,60.00){\line(0,1){40.00}}
\put(82.00,70.00){\vector(0,1){2.00}}
\put(82.00,85.00){\vector(0,1){3.00}}
\put(78.00,100.00){\line(0,-1){40.00}}
\put(78.00,89.00){\vector(0,-1){2.00}}
\put(78.00,74.00){\vector(0,-1){3.00}}
\put(75.50,80.00){\oval(5.00,4.00)[r]}
\put(83.50,80.00){\oval(3.00,4.00)[l]}
\put(69.00,82.00){\vector(-1,0){3.00}}
\put(64.00,78.00){\vector(1,0){3.00}}
\put(106.00,102.00){\circle{4.47}}
\put(106.00,102.00){\makebox(0,0)[cc]{z}}
\put(101.00,85.00){\makebox(0,0)[cc]{$C_+$}}
\put(63.00,84.00){\makebox(0,0)[cc]{$C_-$}}
\put(80.00,55.00){\makebox(0,0)[cc]{$\vdots$}}
\put(80.00,107.00){\makebox(0,0)[cc]{$\vdots$}}
\put(84.00,66.00){\makebox(0,0)[lt]{$C_0$}}
\put(76.00,66.00){\makebox(0,0)[rt]{$C_0$}}
\put(53.00,82.00){\line(1,0){23.00}}
\put(53.00,78.00){\line(1,0){23.00}}
\put(94.00,82.00){\vector(1,0){2.00}}
\put(97.00,78.00){\vector(-1,0){2.00}}
\put(83.00,82.00){\line(1,0){24.00}}
\put(83.00,78.00){\line(1,0){24.00}}
\end{picture}
\caption{\label{Fig:Matsu} The set of contours for the imaginary-time
approach to thermodynamics using thermal field theory. Contour $C_0$
encloses the Matsubara poles of the integrand. Contours $C_+$ and
$C_-$ are the ones needed in a real time formulation of the thermal
field theory in the 4 component representation for fermion fields.}
\end{figure}
Since the integrand of the $z$ integration approaches to zero fast
enough in the $|z|\rightarrow\infty$ limit and is analytic in the
complex $z$ plane excluding the real and imaginary axis, the
integration over $z$ along contour $C_0$ is equivalent to the sum of
integration along contours $C_+$ and the negative of $C_-$ of
Fig. \ref{Fig:Matsu}. The internal energy $U$, entropy $S$ and the
particle number $N$ of the system extracted from $\Omega$ 
according to pattern Eq. \ref{therm1} are
\begin{eqnarray}
 U &=& \frac V{\pi^2}\int_0^\infty d\mbox{\boldmath $p$}
       \mbox{\boldmath $p$}^2 \left [f^{(-)}_\mathbf{p} + f^{(+)}_\mathbf{p}\right ] E_\mathbf{p},\\
 S &=& S^{(+)} + S^{(-)},\\
 S^{(\pm)} &=& - \frac V{\pi^2}\int_0^\infty d\mbox{\boldmath $p$}
       \mbox{\boldmath $p$}^2 \left [f^{(\pm)}_\mathbf{p}\ln f^{(\pm)}_\mathbf{p} + 
      \left (1-f^{(\pm)}_\mathbf{p} \right )\ln \left 
     (1-f^{(\pm)}_\mathbf{p} \right ) \right ],\\
 N  &=&  \frac V {\pi^2}\int^\infty_0 d\mbox{\boldmath $p$}
                  \mbox{\boldmath $p$}^2
                 \left [f^{(-)}_\mathbf{p} - f^{(+)}_\mathbf{p}\right ]
\end{eqnarray} 
with $E_\mathbf{p}=\sqrt{\mbox{\boldmath $p$}^2+m^2}$ and
\begin{equation}
f^{(\pm)}_\mathbf{p} =  \frac1 {e^{\beta(E_\mathbf{p}\pm \mu)}+1} 
\end{equation}
the Fermi--Dirac distribution for fermionic particles and
antiparticles respectively.  These quantities are in agreement with
elementary statistical mechanics.

\subsection{Real-time formulation of the thermal field theory}
\label{sec:RTF}

One of the non-perturbative treatments of the relativistic quantum
field theory is based upon the FMS path integration in real-time
\cite{FMS}. The FMS formalism expresses time evolution between initial
and final states in terms of an integration over paths that connect
the initial and final states with a weight determined by the action of
each particular path. The results of the path integration at the
formal level are not uniquely defined in the Minkowski space-time due
to the presence of singularities.  Their uniqueness is determined by
the fact that the derivation of the FMS path integration
representation of the evolution operator implies a particular ordering
of the intermediate states which defines how the singularities of the
formal results are handled.  The FMS causal structure, which provides
one of the Lorentz invariant prescriptions, determines the particle
content of the quantum fields. It requires that particles are positive
energy solution of the Dirac equation (or corresponding
non-relativistic equation in condensed matter systems with its energy
defined relative to the Fermi surface of the system) that propagate
forward in time and antiparticles (or holes) are the negative energy
solution that effective ``propagate backward in time''. In fact the
canonical quantization of the quantum field is possible only after
such a classification of the solution of the classical ``wave
equation''.  The great success of the quantum field theory in such a
causal structure for real-time processes at zero density 
leaves almost no room for any alternation of it.  Thus the path
integration formalism is defined by not only the formal expressions
that contain singularities but also by the ``causal structure''
following the time ordering of the physical intermediate states.

The dynamical evolution of a system in more general situations
including the finite density ones has also to be studied in such a
real-time formulation of the quantum field theory.  Consistency
requires that the results of such a formulation to agree with the
results from the imaginary-time formulation when it is applied to the
equilibrium situations.  The question is whether or not it actually
happens, especially for fermions. 

\subsubsection{Fermion propagator in finite temperature theory}

 For the fermions interested in this study, the boundary condition
Eq. \ref{Fermion_BC} can be expressed as $\widehat\Psi(t) =
-e^{\beta\widehat K}\widehat\Psi(t) e^{-\beta\widehat K}$. It is
equivalent to Eq. \ref{Fermion_BC} due to the fact that the conserved
$\widehat N$ commutes with the total Hamiltonian $\widehat H$, which
allows the factorization of the action of $\beta\mu\widehat N$ and
$\beta\widehat H$ in the exponential of $\beta\widehat K$. The
$\exp(\beta\mu O_3)$ factor in Eq.  \ref{Fermion_BC} is the result of
the action of $\exp(\beta\mu\widehat N)$ and $\exp(-\beta\mu\widehat
N)$ on both side of $\Psi$. The commutativity of $\widehat H$ and
$\widehat N$ shall not be imposed at this level of development but
rather at later stages as dynamical constraints.

The real-time thermal field dynamics in the 8-component representation
for fermion fields can be constructed along the same line as that of,
e.g., Refs. \cite{LandsWeer,Kapusta,RTTFD1,RTTFD2,RTTFD3,Shuyak}.  The
formal differences of the present approach against those developments
are 1) the Kubo--Martin--Schwinger (KMS) boundary condition for the
contour propagator do not contain the $\exp(\beta\mu)$ factor in the
present approach, the effects of $\mu$ is hidden in the energy
variable within the propagator (namely, the time evolution is
generated by $\widehat K$ not $\widehat H$) here, 2) the Feynman rules
in a perturbation expansion is different from the 4-component theory
and 3) the analytic properties of the present approach in the complex
energy plane is different from some of the earlier ones of this field.

The matrix form of contour propagator for a fermion is
\begin{eqnarray}
  {\cal S}(p) & = & M \left ( \begin{array}{cc}
                   S^{+}(p) &0 \\ 0& S^{-}(p)
                   \end{array} \right ) M,
\label{Mat-Prop}
\end{eqnarray}
where the retarded and advanced propagators $S^+$ and $S^-$ are 
\begin{equation}
S^\pm(p) = \frac
{\pm i} { \rlap\slash p + \gamma^0 \mu O_3 - m O_3 \pm i \gamma^0\epsilon }
\label{S8pm}
\end{equation}
 and
\begin{equation}
M=\left ( \begin{array}{cc} \sqrt{1-n(p_0)} & -\sqrt{n(p_0)}\\
         & \\
                   \sqrt{n(p_0)} & \sqrt{1-n(p_0)} \end{array}
       \right )
\label{TransM}
\end{equation}
with 
\begin{equation}
n(e) =\frac 1 {e^{\beta e}+1}.
\end{equation}
It is easy to show that ${\cal S}_{11}(p)\to S_F(p)$ in the zero
temperature limit since the energy ($p_0$) poles of it are located
above the contour $C_{FMS}$ if they are negative and below it
otherwise. It can be compared to the conventional real-time approach
with the correct causal structure \cite{Shuyak,Lutz} at zero
temperature, in which the poles of the fermion propagator are located
above the $p_0$ integration contour if they are on the left of $\mu$
or else below it.

There is at least another way to realize the KMS boundary condition.
The choice of Feynman propagators and its complex conjugation is
adopted instead of the retarded and advanced ones to express ${\cal
S}(p)$ in some of the earlier literatures of the field. In such a
choice, the matrix $M$ as a function of $p^0$ can not be analytically
continued into the complex $p^0$ plane. It therefore has bad analytic
properties.

The transformation matrix $M$ (Eq. \ref{TransM}) is determined by
the KMS boundary condition. It is therefore independent of the
detailed dynamics of the system and satisfies
\begin{equation}
   M\eta M = \eta,
\label{Mproperty}
\end{equation}
where the matrix
\begin{equation}
  \eta=\left ( \begin{array}{cc} 1 & 0 \\ 0 & -1 \end{array} \right )
\end{equation}
is a matrix acting on the same space as $M$ does.

\subsubsection{The effective or grand potential}

 The grand potential or the effective potential of a free fermionic 
system in the present approach is
\begin{equation}
\Omega = -\lim_{\Delta t\rightarrow\infty} \frac i {2 \Delta t} 
            \mbox{Sp}\left [\mbox{Ln}\gamma^0\eta {\cal S}\right ]_{11} 
             - \Delta\Omega,
\label{Omegax}
\end{equation}
where Sp denotes the functional trace, $\Delta t=t_f-t_i$ (see
Fig. \ref{fig:RTcnt}) and $\Delta\Omega$ is a constant that makes
$\Omega$ to vanish in the zero temperature and density limit. The
subscript ``11'' means that the first diagonal matrix element of the
operator in the square bracket of Eq. \ref{Omegax} within the $2\times
2$ space of the real-time theory for finite temperature should be
taken. Using the propagator Eq. \ref{Mat-Prop} and the property
Eq. \ref{Mproperty} for the matrix $M$, it is easy to show that the
functional trace in this case can be carried out in the 4-momentum
space and
\begin{equation}
 \Omega = -\frac i 2 V \int \frac {d^3 p}{(2\pi)^3} \int\frac
{dp_0}{2\pi} \mbox{Tr} \left [M^2_{11} \mbox{Ln} \gamma^0 S^+(p) -
M_{12}M_{21} \mbox{Ln} \gamma^0 S^-(p) \right ] - \Delta\Omega.
\label{RTgrandP}
\end{equation}
For a free theory the retarded/advanced propagators $S^\pm(p)$ are
given in Eq.  \ref{S8pm}. The resulting $\Omega$ is
\begin{eqnarray}
 \Omega &=& iV \int
\frac {d^3 p}{(2\pi)^3} \left \{ \int_{C_R} \frac {dp_0}{2\pi} \left
[1-n(p_0) \right ]\ln (p_+^2-m^2)(p_-^2-m^2 ) \right . \nonumber \\ &&
\hspace{2cm} + \left .  \int_{C_A} \frac{dp_0}{2\pi} n(p_0) \ln
(p_+^2-m^2)(p_-^2-m^2) - 2 \int_{C_{FMS}}\frac {dp_0}{2\pi}
    \ln (p^2-m^2)  \right \},
\label{RTOmega}
\end{eqnarray}
with $p_\pm^\mu=\{p^0\pm\mu,\mbox{\boldmath{$p$}} \}$, the contour
$C_R$ running above the real $p_0$ axis and contour $C_A$ running
below the real $p_0$ axis. Since the logarithmic functions in the
above equation is well behaved at large $|p_0|$ in the complex $p_0$
plane for the 8-component theory, as it is discussed in detail in the
following, the integration contour for $p_0$ can be deformed to $C_0$
of Fig. \ref{Fig:Matsu}. It can be shown that Eq. \ref{RTOmega} is
exactly the same as Eq. \ref{EuclActC0}. Some of the technical points
are discussed in the following.

Thus the real-time approach using an 8-component ``real''
representation for fermions is equivalent to the conventional
imaginary-time theory. Therefore it give us correct thermodynamics
without additional subtraction terms. Such an equivalence is not
present in the conventional real-time formulation of the thermal field
theory in the 4-component representation for fermions. It is
demonstrated in the following.

\subsection{The zero temperature limit of the real-time theory}

The difference between the 4-component theory and the 8-component one
for fermions manifest in the zero temperature limit already.

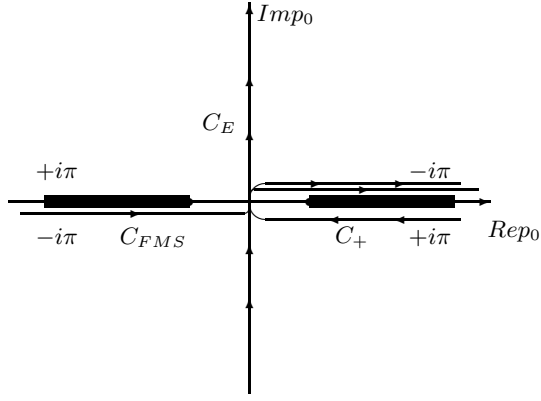
\begin{figure}[h]
\unitlength=0.80mm
\linethickness{0.5pt}
\begin{picture}(120.00,113.00)(20,10)
\put(40.00,80.00){\vector(1,0){80.00}}
\put(124.00,75.00){\makebox(0,0)[cc]{$Rep_0$}}
\put(86.00,111.00){\makebox(0,0)[cc]{$Imp_0$}}
\put(90.00,78.95){\rule{24.00\unitlength}{2.10\unitlength}}
\put(46.00,78.95){\rule{24.00\unitlength}{2.10\unitlength}}
\put(90.00,80.00){\circle*{1.50}}
\put(70.00,80.00){\circle*{1.50}}
\put(42.00,78.00){\line(1,0){36.00}}
\put(78.00,79.00){\oval(4.00,2.00)[rb]}
\put(82.50,81.00){\oval(5.00,2.00)[lt]}
\put(82.00,82.00){\line(1,0){36.00}}
\put(96.00,82.00){\vector(1,0){4.00}}
\put(57.00,78.00){\vector(1,0){5.00}}
\put(80.00,48.00){\vector(0,1){65.00}}
\put(80.00,62.00){\vector(0,1){2.00}}
\put(80.00,71.00){\vector(0,1){2.00}}
\put(80.00,90.00){\vector(0,1){2.00}}
\put(80.00,98.00){\vector(0,1){3.00}}
\put(83.00,83.00){\line(1,0){32.00}}
\put(83.00,77.00){\line(1,0){32.00}}
\put(83.50,80.00){\oval(7.00,6.00)[l]}
\put(90.00,83.00){\vector(1,0){2.00}}
\put(104.00,83.00){\vector(1,0){2.00}}
\put(106.00,77.00){\vector(-1,0){2.00}}
\put(95.00,77.00){\vector(-1,0){2.00}}
\put(64.00,74.00){\makebox(0,0)[cc]{$C_{FMS}$}}
\put(75.00,93.00){\makebox(0,0)[cc]{$C_E$}}
\put(97.00,74.00){\makebox(0,0)[cc]{$C_+$}}
\put(110.00,85.00){\makebox(0,0)[cc]{$-i\pi$}}
\put(110.00,74.00){\makebox(0,0)[cc]{$+i\pi$}}
\put(48.00,74.00){\makebox(0,0)[cc]{$-i\pi$}}
\put(48.00,85.00){\makebox(0,0)[cc]{$+i\pi$}}
\end{picture}
\caption{\label{Fig:Cont1} The set of contours belonging to the same
  FMS class. Contour $C_{FMS}$ is
  the original FMS contour in the Minkowski space. Contour $C_E$ is
  the Euclidean contour. Contour $C_+$ is the quasiparticle contour.
   $\pm i\pi$ denote the imaginary part of the integrand along the
   edges of its cuts (thick lines) on the physical $p_0$ plane.}
\end{figure}

The 4-component theory for the grand potential in the real-time
formulation of the finite density problem at zero density is given
\cite{ThPap,ThPap2} by
\begin{eqnarray}
   \Omega &=& \lim_{\Delta t\rightarrow\infty} \frac i{\Delta t} 
      \mbox{Sp}\mbox{Ln}\gamma^0
       \left ( i\rlap\slash\partial + \mu \gamma^0-m \right ) + \mbox{const}\\
          &=& 2 i V\int \frac {d^3p}{(2\pi)^3}\int_{C_{FMS}} 
        \frac{dp^0}{2\pi} \ln \frac {p_+^2-m^2}{p^2-m^2},
\label{MinkAct}
\end{eqnarray}
where $C_{FMS}$ is the $p^0$ integration contour for the FMS causal
structure shown in Fig. \ref{Fig:Cont1}. 

There are also a pair of contours $C_+$ and $C_-$ in
Fig. \ref{Fig:Matsu}, called the quasiparticle contour \cite{ThPap},
that contribute to the real-time response of the system in the
$\beta\rightarrow\infty$ limit. Contours $C_{FMS}$, $C_E$, $C_+$ and
$C_-$ in Fig. \ref{Fig:Matsu} belong to the same topological class of
contours having the same FMS causal structure. The integration on
$C_E$ leads to the correct thermodynamics as shown in the above
subsection.  Consistency requires the equivalence of the set of
contours $C_{FMS}$, $C_E$, $C_+$ and $C_-$ for the physical
quantities. Eq. \ref{MinkAct} unfortunately fails to meet this
requirement due to the fact that the imaginary part of the logarithmic
function falls off as $O(\mu/|p_0|)$ on the physical $p_0$ sheet.
This makes the results obtained by doing the $p_0$ integration on the
above mentioned set of contours different from each other since the
integration on the large circle sections of the contour has a
non-vanishing value. It is not difficult to verify that while
integrating along $C_E$ produces the correct thermodynamics, explicit
computation of Eq. \ref{MinkAct} on contour $C_+$ or $C_-$ on which
only the imaginary part of the integrand contributes (see Fig.
\ref{Fig:Cont1}), yields a form differs from the finite value $\Omega=
U - \mu N$. It actually diverges.

The origin of the above mentioned problems can be traced back to the
asymmetric nature of the 4-component representation of the fermion
fields with respect to particles and antiparticles in the Minkowski
space-time. If one compares Eq. \ref{EuclActC0} in the
$\beta\rightarrow \infty$ limit and Eq. \ref{MinkAct}, after a
distortion of contour $C_0$ to $C_+$ and the negative of $C_-$ in
Fig. \ref{Fig:Matsu}, one finds that Eq. \ref{MinkAct} differs from
Eq. \ref{EuclActC0} by the lack of the contribution from the
integration along the negative of contour $C_-$, which has to be
present for the finiteness of the result.

However the integration contour along the negative of $C_-$ is absent
for the real-time dynamical evolution of the system according to the
FMS causal structure. There seems to be no way of getting a finite
result that agrees with thermodynamics in the 4-component theory
without at lease one arbitrary subtraction in addition to the zero
density and temperature ones.

The large energy behavior of the 8-component theory for fermions is
better. The grand potential at zero temperature in the 8-component
theory is given by the $\beta\to\infty$ limit of Eq. \ref{RTOmega},
which is
\begin{eqnarray}
  \Omega &=& i V
  \int \frac {d^4 p} {(2\pi)^4} \ln\frac {(p_+^2 - m^2)(p_-^2 - m^2)}
    {(p^2 - m^2)^2}
\label{EuclAct4}
\end{eqnarray}
with the same order of integration as Eq. \ref{MinkAct}. Here
$p_\pm^\mu=(p^0\pm\mu,\mbox{\boldmath $p$})$.  Eq. \ref{EuclAct4} and
Eq.  \ref{MinkAct} have an identical value on the contour $C_E$. They
differ on contours $C_\pm$ because the large $p_0$ behavior of the
imaginary part of the logarithmic function in Eq. \ref{EuclAct4} is of
order $O(\mu^2/|p_0|^2)$ on the physical sheet, which guarantee the
equivalence between the set of contours $C_{FMS}$, $C_E$, $C_+$ and
$C_-$. By following the $C_+$ contour shown in Fig. \ref{Fig:Cont1},
it is simple to show that the resulting right hand side of
Eq. \ref{EuclAct4} is finite and unique, namely, $U - \mu N$. It is
the zero temperature grand potential for a free fermion system at
density $\overline n= N/V$ expected from thermodynamics.

In the 8-component theory for fermions, the above mentioned effects of
the $C_-$ are provided by the lower 4 component $\psi_2$ of $\Psi$.

\subsection{Lorentz covariance and ground state expectation of 
           local observables}
\label{sec:relativity}

 Let us turn to the study of the ground state expectation of 
local bilinear operators constructed from
two fermion fields of the form 
\begin{equation}
\widehat O = \overline\Psi(x)\Gamma\Psi(x)
\end{equation}
in a finite density environment where $\Gamma$ is certain matrix acting
on $\Psi$.  A local product of two field operators is in general
singular and non-unique. The usual procedure of normal ordering
depends on the Fock space in which the particles of the corresponding
``non-interacting'' Lagrangian is represented. In strong interaction,
these particles, like the current quarks, may never appear in the
asymptotic spectra of the full theory. In addition, there is a
non-perturbative spontaneous chiral symmetry breaking in the vacuum
state, which generates a much larger mass for the quarks. It makes a
unique definition of the normal ordering for strong interaction
theories quite difficult.

A definition of such a potentially divergent product that is
independent of the dynamics of the system is 
\begin{equation}
 \widehat O = \lim_{\delta_\mu\rightarrow 0} 
                       \overline\Psi(x+\delta)\Gamma\Psi(x), 
\end{equation}
where $\delta_\mu$ is a 4-vector with, e.g., $\delta_0<0$.  The
ground state (vacuum) expectation value of $\widehat O$ can thus be
computed from the fermion propagator ${\cal S}$ 
\begin{eqnarray}
   <\widehat O> &=& - \mbox{tr} \int \frac {d^3p} {(2\pi)^3}\int_{C_R}
                        \frac {dp_0} {2\pi} {\cal S}^{11}(p)\Gamma,
\label{avO}
\end{eqnarray}
with ``tr'' denoting the trace over internal indices of the fermion
fields and
\begin{equation}
 {\cal S}^{11} = \left [1-n(p^0)\right ] S^+(p) - n(p^0) S^-(p).
\end{equation}
Since there is no poles and cuts off the real $p_0$ axis except the
ones on the imaginary $p_0$ axis due to the thermal factor, which
should be excluded (see Fig. \ref{Fig:Matsu}), contour $C_R$ can be
closed in the lower half plane to include the poles of the
retarded propagator $S^+$.
So, Eq. \ref{avO} is reduced to
\begin{eqnarray}
  < \widehat O > &=& i\int \frac {d^3p} {(2\pi)^3}\sum_k [1-n(p^0_k)]
                   \mbox{tr}\mbox{Res}
                    S^{+}(p_k)\Gamma,
\label{avO1}
\end{eqnarray}
where the sum is over all poles $p^0_k$ of $S^{+}(p)$ and ``Res''
denotes the corresponding residue.  The difference between the
8-component theory and the 4-component one also manifests here. For
example, the conserved fermion number density $\overline n$ corresponding
to current
\begin{equation}
j^\mu = V_I^\mu = 
\frac 1 2 \overline\Psi \gamma^\mu O_3 \Psi
\end{equation}
 is
\begin{equation}
       \overline n  =  {1\over\pi^2}\int^\infty_0 d\mbox{\boldmath $p$}
                  \mbox{\boldmath $p$}^2
                 \left [f^{(-)}_\mathbf{p} - f^{(+)}_\mathbf{p}\right ].
\label{av01}
\end{equation}
 It is same as the
one in elementary statistical mechanics.

If we choose $\delta_0>0$,
then the contributing component of ${\cal S}^{11}$ is the advanced propagator
$S^{-}$
\begin{eqnarray}
  < \widehat O > &=& -i\int \frac {d^3p} {(2\pi)^3}\sum_k n(p^0_k)
                   \mbox{tr}\mbox{Res}
                    S^{-}(p_k)\Gamma.
\label{av02}
\end{eqnarray}
It is the same as Eq. \ref{avO1}. It also tells us that the choice
made in Eq. \ref{intLag} is the right one. This is because if we
choose ``type--II'' operators for the fermion number density instead,
the result is divergent.

On the other hand, $\overline n$ computed in the 4-component theory
using such a point split definition of fermion number density is
divergent and different from each other for $\delta_0>0$ and
$\delta_0<0$. It means that it is not even consistent with relativity
for a space like $\delta_\mu$ whose time component can has different
sign in different reference frames. It can be made finite and unique
only after an arbitrary subtraction that depends on the reference
frame. It is another serious conceptual problem originated from the
non-commutativity of quantum mechanical quantities and the
relativistic space-time. Such a conflict between quantum mechanics and
relativity is avoidable only in the 8-component theory for fermions.

\section{Quantization and Particle Interpretation}
\label{sec:Particle}

 The language used so far is the path integration one which treats the
fermion fields as Grassmanian numbers. Such a representation of the
problem is most useful in discussing the wave aspects of the
problem. In most experimental observations, especially in high energy
physics in which only a small number of particles are involved in a
reaction, the particle aspect of the problem has to be understood. In
such a case, the physics is most easily described in terms of
the Fock space of the problem in the operator language.

\subsection{Free particles}

 The particle content of the 8-component theory can be found by first
studying the pole structure of the fermion propagator under the FMS
causal condition. The time dependence of the propagator (see section
\ref{sec:FDQFT}) at zero temperature is
\begin{equation}
\lim_{\beta\rightarrow\infty}{\cal S}_{11}(t,\mbox{\boldmath $p$}) 
= S_F(t,\mbox{\boldmath $p$})  =\int_{C_{FMS}} \frac {dp_0} {2\pi} 
e^{-ip_0 t} S_F(p_0,\mbox{\boldmath $p$}).
\label{SFt}
\end{equation}

In case of $t>0$, Eq. \ref{SFt} can be evaluated on the contour $C_+$
of Fig. \ref{Fig:Matsu} by closing the integration contour in the
lower half $p_0$ plane
\begin{eqnarray}
    S_F(t,\mbox{\boldmath $p$}) 
               &=& \theta(E_\mathbf{p}-\mu) \Lambda^{1+}_\mathbf{p} e^{-i(E_\mathbf{p}-\mu)t} +
               \theta(\mu-E_\mathbf{p})   
                \Lambda^{2-}_\mathbf{p}
                 e^{-i(\mu-E_\mathbf{p})t} + 
               \Lambda^{2+}_\mathbf{p} e^{-i(E_\mathbf{p}+\mu)t}  
\label{PropTgt0}
\end{eqnarray}
and, if $t<0$, can be evaluated on the contour $C_-$ to obtain
\begin{eqnarray}
   S_F(t,\mbox{\boldmath $p$}) 
               &=& \Lambda^{1-}_\mathbf{p}  e^{i(E_\mathbf{p}+\mu)t} +
               \theta(\mu-E_\mathbf{p}) \Lambda^{1+}_\mathbf{p} e^{i(\mu-E_\mathbf{p})t} + \theta(E_\mathbf{p}-\mu) 
               \Lambda^{2-}_\mathbf{p} e^{i(E_\mathbf{p}-\mu)t}.    
\label{PropTlt0}
\end{eqnarray}
Here 
\begin{equation}
\Lambda_\mathbf{p}^{r\pm} = P_r\frac {(\pm\gamma^0 E_\mathbf{p} - \mbox{\boldmath
  $\gamma$}\cdot \mbox{\boldmath $p$} + m O_3)} {2E_\mathbf{p}}, 
\end{equation}
$P_1=(1+O_3)/2$, $P_2=(1-O_3)/2$ are projection operators and $r=\{1,2\}$.

\begin{figure}[ht]
\includegraphics{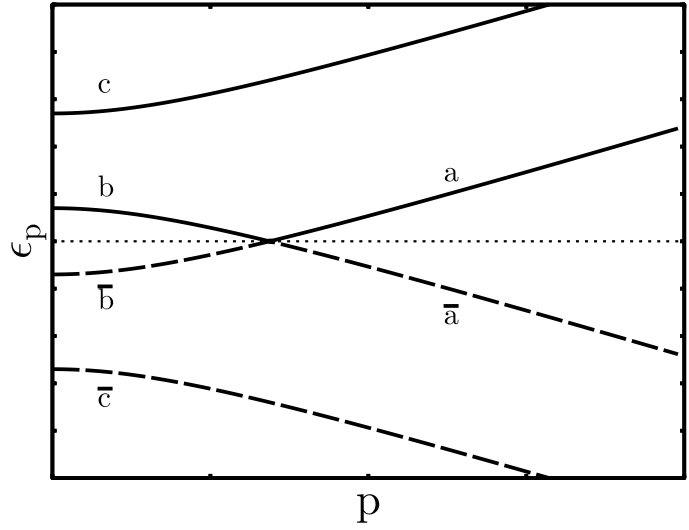}
\caption{\label{Fig:spectra} A schematic plot of the spectra for fermions
in the 8-component theory when $\mu>m$. The solid lines above zero (of
$p^0$ in the propagator) correspond to particles and the dashed line
below zero correspond antiparticles. The $p^0=0$ surface correspond to
the Fermi surface.}
\end{figure}

The FMS causal structure in the present theory can be simply putted
as: 1) excitations with $p_0> 0$ that propagate {\em forward in time}
correspond to particles and 2) those with $p_0< 0$ that propagate {\em
backward in time} correspond to antiparticles. Fig. \ref{Fig:spectra}
shows the spectra for particles and antiparticles at finite density
with $\mu>m$, where lines $a$, $\overline b$ and $\overline c$ are
contributions from the $r=1$ excitations and lines $\overline a$, $b$
and $c$ are contributions from the $r=2$ ones.

The quantization of the 8-component fermionic field then follows
naturally. $\widehat \Psi(x)$ can be written as
\begin{eqnarray}
    \widehat \Psi(x) &=& \sum_{r\mathbf{p}s} \frac 1 {2 E_\mathbf{p}}
       \left [ U_{r\mathbf{p}s} e^{-i\theta({\bf p})\widetilde p\cdot x} \widehat B_{r\mathbf{p}s}(t) +
           V_{r\mathbf{p}s} e^{i\theta({\bf p})\widetilde 
                 p\cdot x} \widehat D^\dagger_{r\mathbf{p}s}(t)\right ],
\label{PsiOp}
\end{eqnarray}
where when $\mu>m$, 
\begin{eqnarray}
      \widehat B_{1\mathbf{p}s}(t) = \left \{ \begin{array}{cr}
                          e^{i\mu t} \widehat b_{1\mathbf{p}s} & \hspace{0.3cm}
                            \mbox{if $|{\bf p}| > p_F $}\\
                          e^{-i\mu t} \widehat b_{1\mathbf{p}s}^\dagger &
                          \hspace{0.3cm}
                          \mbox{if $|{\bf p}| < p_F $}
                         \end{array} \right .
         &\hspace{0.3cm}\mbox{and}\hspace{0.3cm}&
      \widehat B_{2\mathbf{p}s}(t) = e^{-i\mu t} \widehat b_{2\mathbf{p}s} \\
      \widehat D^\dagger_{2\mathbf{p}s}(t) = \left \{ \begin{array}{cr}
                          e^{i\mu t} \widehat d^\dagger_{2\mathbf{p}s} & \hspace{0.3cm}
                            \mbox{if $|{\bf p}| > p_F $}\\
                          e^{-i\mu t} \widehat d_{2\mathbf{p}s} &
                          \hspace{0.3cm}
                          \mbox{if $|{\bf p}| < p_F $}
                         \end{array} \right .
         &\hspace{0.3cm}\mbox{and}\hspace{0.3cm}&
      \widehat D^\dagger_{1\mathbf{p}s}(t) = e^{-i\mu t} \widehat d^\dagger_{1\mathbf{p}s}
\end{eqnarray}
with $p_F=\sqrt{\mu^2-m^2}$ and $\theta({\bf p})=\mbox{sign} (|{\bf
p}|-p_F)$.  Here $s$ is the spin index, the 4-momentum $\widetilde
p^\mu = \{E_{\bf p},{\bf p}\}$, $U_{r\mathbf{p}s}$ and
$V_{r\mathbf{p}s}$ are 8-component spinors that satisfy
\begin{eqnarray}
(\rlap\slash {\widetilde p} - mO_3)U_{r\mathbf{p}s}&=&0, \\
(\rlap\slash {\widetilde p} + mO_3)V_{r\mathbf{p}s}&=&0.
\end{eqnarray}
Their solution can be easily found
\begin{eqnarray}
    U_{1\mathbf{p}s} = \left ( \begin{array}{c} u(\mathbf{p}s) \\ 0 \end{array} \right ),
    &\hspace{0.4cm} &
    U_{2\mathbf{p}s} = \left ( \begin{array}{c} 0 \\ 
        v(\mathbf{p}s) \end{array} \right ), \\
    V_{1\mathbf{p}s} = \left ( \begin{array}{c} v(\mathbf{p}s) \\ 0 
                   \end{array} \right ),
    &\hspace{0.4cm} &
    V_{2\mathbf{p}s} = \left ( \begin{array}{c} 0 \\ u(\mathbf{p}s) 
                  \end{array} \right )
\end{eqnarray}
with $u(\mathbf{p}s)$ and $v(\mathbf{p}s)$ the conventional Dirac
spinors for particles and antiparticles respectively in the
4-component theory. It can be seen that the two $U$s with the same
quantum numbers besides $r$ map into each other under the mirror
reflection transformation so do the two $V$s.

The non-vanishing anticommutators between
$\widehat b_{r\mathbf{p}s}$ and $\widehat d_{r\mathbf{p}s}$ 
in the canonical quantization of the fermion fields are
\begin{eqnarray}
   \left \{\widehat b_{r\mathbf{p}s},\widehat b^\dagger_{r'\mathbf{p}'s'} 
     \right \} &=&
        2E_\mathbf{p}\delta_{\mathbf{pp}'}\delta_{ss'}\delta_{rr'}, \\
   \left \{\widehat d_{r\mathbf{p}s},\widehat d^\dagger_{r'\mathbf{p}'s'} 
     \right \} &=&
        2E_\mathbf{p}\delta_{\mathbf{pp}'}\delta_{ss'}\delta_{rr'}.
\end{eqnarray}
They realize a canonical quantization of $\widehat \Psi$. 

\subsection{The constraint of the reality condition}
\label{sec:qrealcond}

  Since the mirror reflection operations defined in Eqs. \ref{M-def}
and \ref{RM-def} involve time reversal transformation, it is necessary
to discuss the time reversal transformation on operators in the
Hilbert space in some details first.

\subsubsection{The causal and motion reversal of time}
\label{sec:qTRVT}
  There are two kinds of statement for time reversal transformation
of the matrix elements \cite{thesis} of a operator. 
The first one is the so called causal reversal transformation which involves
the interchange of initial and final states together with the motion reversal
of the quantum numbers, like the 3--momentum, $z$--component of the
angular momentum, etc., of the initial and the final states. Namely
\begin{equation}
   <f|\widehat O |i> \stackrel{\widehat {\cal T}_{c}}\longrightarrow 
          <\widetilde i|\widehat O
      |\widetilde f>
\label{c-TR}
\end{equation}
with the tilde states the corresponding motion reversed one. The second one
is the so called motion reversal transformation which is defined by
\begin{equation}
   <f|\widehat O |i> \stackrel{\widehat{\cal T}_{m}}\longrightarrow 
        <\widetilde f|\widehat O
      |\widetilde i>^*,
\label{m-TR}
\end{equation}
which contains a complex conjugation but does not exchange the initial
and final states. These two definitions are equivalent only for
Hermitian operators.

The genuine test of time reversal invariance
involves causal reversal of time \cite{thesis,cern1} because 
it can be shown that the signals in the 
motion reversal test of time reversal invariance could be contaminated
by pseudo time reversal violation effects \cite{thesis,EMH-ME} due
to final state interaction. 

The operator form of causal reversal of time given
by Eq. \ref{c-TR} is 
\begin{equation}
   \widehat {\cal T}_c \widehat O \widehat {\cal T}^{-1}_c =
       \widehat U \widehat O^T \widehat U^\dagger
\label{op-c-TR}
\end{equation}
with superscript ``T'' denoting the transpose of the operator and
$\widehat U$ an unitary operator that maps a state $|\phi>$ to its
motion reversed state $|\widetilde\phi>$. 

      For the fermion field operator $\widehat\Psi$, the operator
causal time reversal transformation corresponding to Eq. \ref{T-rev-8} is
\begin{equation}
    \widehat \Psi({\bf x},t) \stackrel{\widehat{\cal T}}{\longrightarrow}
              \gamma^0\gamma^5 iO_2 \widehat \Psi^T({\bf x},-t),
\label{T-rev-8-op}
\end{equation}
where the transpose on the right hand side is only on the operators $
\widehat b$s and $\widehat d$s but not on the Dirac indices. From
Eq. \ref{PsiOp}, taking into account that the time reversal
transformation is an anti-linear transformation, Eq. \ref{T-rev-8-op}
is equivalent to
\begin{eqnarray}
   \widehat b_{1\mathbf{p}s} \stackrel{\widehat{\cal T}}\longrightarrow \widehat 
                    b_{2-\mathbf{p}-s}^T,
   &\hspace{1cm} & 
   \widehat b_{2\mathbf{p}s} \stackrel{\widehat{\cal T}}\longrightarrow \widehat 
                    b_{1-\mathbf{p}-s}^T,\\
   \widehat d_{1\mathbf{p}s} \stackrel{\widehat{\cal T}}\longrightarrow \widehat 
                    d_{2-\mathbf{p}-s}^T,
   &\hspace{1cm} & 
   \widehat d_{2\mathbf{p}s} \stackrel{\widehat{\cal T}}\longrightarrow \widehat 
                    d_{1-\mathbf{p}-s}^T.
\end{eqnarray}

\subsubsection{The constraint of the reality condition}

 The operator representation of the mirror reflection operation
defined in Eq. \ref{M-def} on $\widehat\Psi$ is then realized by
\begin{eqnarray}
   \widehat b_{1\mathbf{p}s} \stackrel{\widehat{\cal M}}\longrightarrow \widehat b_{2\mathbf{p}s}^*,
   &\hspace{1cm} & 
   \widehat b_{2\mathbf{p}s} \stackrel{\widehat{\cal M}}\longrightarrow \widehat b_{1\mathbf{p}s}^*,\\
   \widehat d_{1\mathbf{p}s} \stackrel{\widehat{\cal M}}\longrightarrow 
     \widehat d_{2\mathbf{p}s}^*,
   &\hspace{1cm} & 
   \widehat d_{2\mathbf{p}s} \stackrel{\widehat{\cal M}}\longrightarrow 
     \widehat d_{1\mathbf{p}s}^*.
\end{eqnarray}

So the reality condition at zero density is
\begin{equation}
   \widehat b_{1\mathbf{p}s} = \widehat b^*_{2\mathbf{p}s}, \hspace{1cm} 
   \widehat d_{1\mathbf{p}s} = \widehat d^*_{2\mathbf{p}s}.
\end{equation}
These conditions for the Fock space are almost trivial to implement as
long as the matrix elements of annihilation operator $\widehat b$s and
$\widehat d$s are chosen to be real number and they are renormalized
in the same way in interacting systems.

\subsection{Feynman rules for elementary processes and the reality condition}
\label{sec:FR-q}

   Although the 8-component theory for fermion contains twice as many
degrees of freedom as the 4-component one, it is shown in the
discussion of the previous sections, especially in section
\ref{sec:FDQFT}, all of these excitation modes should be taken into
account in order to obtain correct counting of states due to the $1/2$
factors contained in the free Lagrangian and in the interaction
vertices. 

  The properties of the ground state, like the grand potential, the
fermion number density and the scalar charge density are all computed
using the quark propagator that contains external background fields,
collectively denoted as $f$, like the chemical potential $\mu$ (or the
statistical gauge field in the local theory \cite{ThPap,ThPap2}), the
scalar field $\sigma$, etc.. In fact each set of these background
fields has a corresponding set of positive energy ($p^0$ in the
propagator) and negative energy excitation modes and therefore a
corresponding temporary ground state for the fermion sector of the
system. The properties of the ground state in the background field $f$
for the fermion sector is obtained by summing over the contributions
of individual excitation modes with negative energy. In the true
ground state of the whole system, including the boson sector, the
external field takes certain value $\overline f$, which is
called the classical configuration that stabilizes the system.

  The elementary processes concerns the local excitations of the
system above the ground state (positive $p^0$ events) in the external
field configuration $\overline f$, which is the stable configurations
of the whole system including the external source. They are treated
differently from the ground state properties.  The quantum fluctuation
part of the bosonic fields $\delta f=f-\overline f$ does not enter the
fermion propagator in the perturbative expansion of the reaction
amplitude of elementary processes. The central question here is how to
implement the anti-linear reality condition Eq. \ref{realcond} for
quantized $\delta f$.

  The general one particle spinor satisfies the 8-component Dirac
equation which can be written as Eq. \ref{Spinor-Eq} in general with
$\Sigma$ subject to constraint Eq. \ref{realcond2}. Suppressing the
spin and the ones corresponding to internal symmetries, there are four
kinds of solutions. Two solutions $\Psi_1$ and $\Psi_2$ with positive
energy and corresponding two solutions with negative energy. In the
computation of physical reaction amplitude at zero density, one should
choose one of the solutions from these two, say $\Psi_1$ (or $\Psi_2$,
it does not matter which one is chosen.) the reality condition
generates the other solution by using the mirror transformation
$\widehat{\cal M}$ on $\Psi_1$, which is denoted as $\Psi_{1{\cal
M}}$. At the one particle level, $\Psi_{1{\cal M}}$ is identical to
$\Psi_2$. Therefore it seems that one can simply sum over $\Psi_1$ and
$\Psi_2$ to get the final results. This is not always right due to the
fact that the mirror transformation is an anti-linear transformation
so that although the single particle matrix element $\Psi_{1{\cal M}}$
is identical to $\Psi_2$, they lives in different time zone. Using the
causal time reversal representation of $\widehat {\cal M}$ discussed
above (one can not use the motion reversal representation since the
complex conjugation contained in it can modifies the causal structure
of the fermion propagators leaving the effects of pseudo-time reversal
violation \cite{thesis}), combined with Eqs. \ref{RC-boson1},
\ref{RC-boson2} and \ref{RC-vert}, it can be shown that the
interaction vertices inside of a Feynman graph is obtained from the
interaction terms in Lagrangian density, which is denoted as $\Gamma$,
by doing the following transformation
\begin{equation}
     \widetilde \Gamma = \Theta \Gamma, 
\label{pert-vtx}
\end{equation}
where $\widetilde \Gamma$, $\Gamma$ and $\Theta$ are $2\times 2$ matrices
in the upper and lower 4-component space of $\Psi$ and
\begin{equation}
     \Theta = \left ( \begin{array}{cc} 1 & 0 \\ 0 & \pm (-1)^\nu \end{array}
             \right )
\end{equation}
with $\pm$ sign corresponding to a vertex of ``type--I'' or ``type--II'' 
respectively, 
where $\nu$ is the number of Lorentz indices on the vertices.  Using
$\widetilde\Gamma$ one can use $\Psi_1$ and $\Psi_2$ or the propagator
$S_F$ to count the physical state in a way that respects the reality
condition. 

For situations in which $D=\overline D=0$ in the mass matrix $\Sigma$
(namely, Eq. \ref{nsc-mass}), it can be shown that the value of the
Feynman diagrams are the same as the conventional 4-component theory
when $\mu=0$. For example, the one photon exchange
electron--electron scattering amplitude contains three kinds of terms
in the 8-component theory. They are $e-e$ scattering, $e-\widetilde e$
scattering and $\widetilde e-\widetilde e$ scattering terms, where
$\widetilde e$ is the mirror electron. These terms are all identical and
repulsive. If the reality condition is 
incorrectly imposed or not imposed at all, there can be
$e-\widetilde e$ attraction leading to the bound $e\widetilde e$ pair
that is not observed in Nature.

On may ask why $\Theta$ does not enter the expression for the grand
potential and the ground state expectation value of local observables.
The answer can be found if one carefully study the evaluation of the
ground state expectation value of local observables. Each
configuration of the background fields defines its own sets of
particles and anti-particles since the spectra of $S^{-1}_F$ depend on
it. In the background fields configuration $\overline f +\delta f$
some of the particles of in configuration $\overline f$ (positive
energy solution) become antiparticles (negative solution) and vice
versa. But perturbation theory is based on the particle content of
configuration $\overline f$, $\delta f$ is treated as quantum
fluctuations above the ground state and not included in the
propagator, the factor the $(-1)^\nu$ is there to compensate for that.
If $\delta f$ is included in the propagator in certain
non-perturbative calculations like the lattice simulation, no $\Theta$
factor is needed, but in this case it is the ``ground state'' (in the
presence of the external $\overline f+\delta f$ field) property that
we are talking about then. As a rule, all vertices in a connected
diagram that couple to a propagator of bosons should be written in
terms of Eq. \ref{pert-vtx}, only tadpoles type of vertices that
couple to classical part of the boson fields or the background fields
should use $\Gamma$ directly.

The reason why there is no such a factor in the 4-component theory is
because the 4-component theory is ``worse'' than that by having
divergences discussed in section \ref{sec:FDQFT}. The delicacies
concerning the factor $(-1)^\nu$ needs not to be concerned since there
are arbitrary infinities not present in the 8-component theory to be
subtracted.

\subsection{Elementary processes at finite density}

 As it is mentioned above the 8-component theory is not equivalent to
the 4-component theory at finite density, especially in relativistic
many body systems. The differences can be traced back to the different
behavior of the particles and their mirror partners in the finite density
situations. 

\subsubsection{The mirror particles}

 The positive energy excitation mode of the 4-component theory at
finite density corresponds to the curve ``a'' of
Fig. \ref{Fig:spectra} only. The mirror partner of ``a'' in the
8-component theory are curves labeled by ``b'' and ``c''. Excitation
``c'' is the mirror partner of ``a'' with respect to the Dirac sea
since these two excitations become identical in the $\mu\to 0$
limit. It is separated from ``a'' by an
energy difference $2\mu$ so that it is highly suppressed in
non-relativistic conditions when $\beta m << 1$ with $\beta$ the
inverse temperature and $m$ the mass of the excitation. Excitation
``b'' is the mirror excitation of ``a'' with respect to the Fermi
sea. Due to asymmetric nature of the Fermi sea, excitation ``b'' is
not identical to excitation ``a'' with respect to the Fermi
surface. Only in the vicinity of the Fermi surface, both of them can
be regarded as approximately identical since both of them have a
linear spectra that is proportional to $|{\bf p}|-p_F$ with constant
density of states. Here, $p_F$ is the Fermi momentum.

Besides the excitation spectra and density of states, the mutual
interaction between particles is also different in the 8-component
theory and in the 4-component theory.  In order to find the difference
let us find out the vector and scalar charges of the particles
corresponding to spectra $a$, $b$ and $c$ and the antiparticles
corresponding to spectra $\overline a$, $\overline b$ and $\overline
c$ of Fig. \ref{Fig:spectra}.

The vector and scalar charge of a particle can be computed by taking the 
expectation value of the corresponding current between the zero momentum
state of that particle. The result is given in Table \ref{tab:charge}.
\begin{table}
\caption{\label{tab:charge} The scalar and vector charges of particles
         and antiparticles in the 8-component theory.}
\begin{tabular}{|c|cccccc|}
\colrule
 Excitations  &\phantom{aaaa} 
$a$\phantom{aaaa} & \phantom{aaaa}$b$\phantom{aaaa} 
&\phantom{aaaa} $c$ \phantom{aaaa}&\phantom{aaaa} $\overline a$\phantom{aaaa} & 
\phantom{aaaa} $\overline b$\phantom{aaaa} &\phantom{aaaa} $\overline c
$\phantom{aaaaaa} \\
\colrule
Scalar Exchange & $1$ & $-1$ & $1$ & $-1$ & $1$  & $-1$ \\
Vector Exchange & $1$ & $1$  & $1$ & $-1$ & $-1$ & $-1$ \\ 
\colrule
\end{tabular}
\end{table}

\subsubsection{The particle--particle scattering by exchange of vector bosons}

In the 4-component theory for fermions at finite density, the excitation
of the system corresponding to particles is given by line $a$ of Fig.
\ref{Fig:spectra}. Let us denote the scattering amplitude between two
excitations of ``$a$'' type in the 4-component theory by $T^{(4)}_{aa}$.
The corresponding scattering processes in the 8-component theory is
of three kind: $aa$, $bb$ and $ab$ (ignoring $c$ for the
moment since it lies too high in energy to be significant
in non-relativistic situations), therefore we have the following
correspondence
\begin{equation}
  T^{(4)}_{aa} \longleftrightarrow \frac 14 \left (
         T_{aa}^{(8)} + T_{bb}^{(8)} + T_{ab}^{(8)} + T_{ba}^{(8)}
\right ),
\label{4-8corr}
\end{equation}
where the $1/4$ factor is obtained by extracting the $1/2$ factor for
the vertices of the 8-component theory.  For the interaction induced
by the exchange of vector bosonic particles like the photon, all four
terms on the right hand side of correspondence (\ref{4-8corr}) have
the same sign since excitation ``a'' and ``b'' have the same vector
charge. It can be shown that $T^{(4)}_{aa}$ and various
$T^{(8)}_{aa,bb,ab,ba}$ also have the same form in terms of
4-component Dirac spinors $u(\mathbf{p}s)$.  The only difference is
that excitations ``a'' and ``b'', which lives in different momentum
domain, have different phase space volume.  However, for condensed
matter system like the electron gas in a metal only the excitations
near the Fermi surface is important. In this case the density of
states for both excitation ``a'' and ``b'' can be approximated by the
density of state of the system at the Fermi surface, and
Eq. \ref{4-8corr} can be viewed as an approximate equation. So the
non-relativistic condensed matter system of the above type in which
the particle--particle interaction is dominated by the vector Coulomb
interaction can not distinguish between the predictions of the
4-component theory and the 8-component theory when the density is
sufficiently high.

\subsubsection{The particle--particle scattering by exchange of scalar bosons}

Since the excitation mode ``b'' has different scalar charge from the
excitation mode ``a'', the predictions of the 8-component theory and
the 4-component theory for fermion--fermion scattering through the
exchange of scalar particles in finite density system is different 
since the prediction of the 8-component theory is
\begin{equation}
      A^{(8)}\sim \frac 14 \left (T^{(8)}_{aa}+T^{(8)}_{bb}
              - T^{(8)}_{ab} - T^{(8)}_{ba} \right ) << T^{(4)}_{aa}
\end{equation}
since any of the $T^{(8)}_{aa,bb,ab,ba}$ inside the bracket and the
corresponding amplitude $T^{(4)}_{aa}$ is of the same sign and order
of magnitude. 

Such a difference can even manifest in non-relativistic condensed
matter systems if their underlying interaction contains scalar type of
vertices.  Although when the non-relativistic reduction is done, the
leading interaction terms contain no information about whether they
are from vector type of interaction or from scalar ones in the
4-component theory, the predictions of the 8-component theory for
these two type of interactions are different even in non-relativistic
situations. The wide success of the 4-component theory in
non-relativistic condensed matter system indicate that a underlying
vector type of interaction dominates the non-relativistic condensed
matter physics according to the 8-component theory for fermions.  In
fact we know it is the quantum electrodynamics that underlies the
non-relativistic condensed matter system.

 The prediction of the 8-component theory for fermions can be tested
in nuclear matter in which there are scalar particles like the
$\sigma$ and $\pi$ mesons that provide the nucleon--nucleon
attraction. The 8-component theory tells us that the effects of
$\sigma$ and $\pi$ mesons beyond the mean field (as it is shown in
\cite{Paps2}, the mean field effects of these scalar fields are the
same as that of the 4-component theory) is much reduced as compared to
the 4-component theory given the same interaction strength in a finite
density environment.  Thus in nuclear matter or heavy nuclei, the
nucleon--nucleon attraction is much reduced leaving the ``residue''
repulsive interaction due to the exchange of the vector mesons. So,
the nucleons in a nuclear matter or a heavy nucleus tend to avoid each
other more than they do in empty space, especially when the density is
high.

\section{Two Issues Concerning Vacuum Color Superconductivity}
\label{app:ColorSuperc}
  Two issues concerning the properties of the possible color
superconducting phase of the ground state of the strong interaction
remain to clarified.  The first one is about the nature of the
competition between the color superconducting phase and the normal
chiral symmetry breaking phase of the strong interaction ground state
as the density of the system is lowered \cite{front}. The second one
is about the nature of the spontaneous $CP$ violation in the color
superconducting phase of strong interaction that is predicted by the
local theory based on the 8-component representation of fermion fields
\cite{Paps1,ThPap}.

\subsection{The grand potential and metastability of the color
            superconducting phase at zero density}

In the Hartree--Fock or mean field approximation, the behavior of the
system can be reasonably approximated by a collection of
quasi-particles with their mutual interaction ignored as a first order
approximation. The effects of the interaction are entirely encoded in
the mass matrix in the propagator for the quasi-particles, which has
the generic expression
\begin{equation}
   S_F(p) = \frac{i}{\rlap\slash p +\gamma^0\mu O_3 - \Sigma + i\epsilon}.
\end{equation}
The mass term is of the following general form
\begin{equation}
   \Sigma = \left ( 
                \begin{array}{cc}
                     \sigma & D \\  \overline D & -\sigma
                \end{array}   
            \right )
\label{qp-mass-mat}
\end{equation}
in order the satisfies the zero density
general reality condition Eq. \ref{realcond2}.

The quantity $\sigma$ is the order parameter for the chiral
symmetry for a massless fermion system. The sub-matrices
$D$ and $\overline D$ is non-vanishing if the system is (color)
superconducting. For example, if the system is in a state of scalar 
color superconducting phase \cite{sca_sc,ThPap}
\begin{equation}
     D= \gamma^5 {\cal A}_c\chi^c, \hspace{0.8cm} \overline D = \gamma^5 
       {\cal A}^c \overline \chi_c
\label{scl-sc-D}
\end{equation}
with $\chi^c$ and $\overline\chi_c$ the pair of order parameters for
the scalar color superconducting phase with
$(\chi^c)^\dagger=-\overline\chi_c$.  If the system is in a state of
vector superconducting phase \cite{Paps1,Paps2}
\begin{equation}
     D= -\phi^c_\mu \gamma^\mu\gamma^5 {\cal A}_c , \hspace{0.8cm} 
     \overline D = \overline{\phi^\mu_c} \gamma_\mu\gamma^5 {\cal A}^c
\end{equation}
with $\phi^c_\mu$ and $\overline{\phi^\mu_c}$ the pair of order parameters
for the vector color superconducting phase.

Take the simpler case of scalar color superconducting phase for
example, the zero temperature and density grand potential for unit
volume (or effective potential) given by Eq. \ref{RTgrandP} for the
quasi-particles with a mass matrix Eq. \ref{qp-mass-mat} for a scalar
color superconducting phase is \cite{sca_sc,ThPap}
\begin{eqnarray}
 \Omega/V &=& 4 i \int \frac{d^4p}{(2\pi)^4} \ln \left [
          \left (1-\frac{\sigma^2+\chi^2}{p^2}   \right )^2 -
         \frac{\sigma^2}{p^2} \left ( 
           1- \frac{\sigma^2-\chi^2}{p^2} \right )^2
\right ] + \frac 1 {4G_0} \sigma^2 + \frac 1 {2G_3} \chi^2\nonumber \\
        &=& 4 i \int \frac{d^4p}{(2\pi)^4} \ln \left \{
\left ( 1-\frac{\sigma^2} {p^2} \right ) 
\left [ \left (1-\frac{\sigma^2+\chi^2} {p^2} \right )^2 - 2\frac{\sigma^2\chi^2}
       {p^4} \right ]
\right \} + \frac 1 {4G_0} \sigma^2 + \frac 1 {2G_3} \chi^2
\end{eqnarray}
with $G_0$ and $G_3$ coupling constants.  

The grand potential at zero temperature and density is a function of
two order parameters: $\sigma$ for the spontaneous chiral symmetry
breaking phase and $\chi$ for the color superconducting phase of the
strong interaction vacuum state. It has two pairs of minima for
sufficiently large $G_0$ and $G_3$. One pair locates at the chiral
symmetry breaking points $\sigma=\pm \sigma_{0}\ne 0$ and $\chi=0$ and
the other pair locates at the color superconducting points $\sigma=0$
and $\chi=\pm\chi_0\ne 0$. There exists a potential barrier between
these points. Phenomenology implies that the present day vacuum state
of the strong interaction is in the chiral symmetry breaking
phase. Therefore the coupling constants $G_0$ and $G_3$ have to be so
chosen that the points $\sigma=\pm\sigma_0\ne 0$ and $\chi=0$ is the
absolute minima of $\Omega$ and the other pair of minima of $\Omega$
are local minima corresponds to a metastable color superconducting
phase at zero density\cite{front}.

It can be shown that in some of the other recent works on color
superconductivity \cite{other}, an equivalent of the following mass
matrix is used
\begin{equation}
   \Sigma = \left ( 
                \begin{array}{cc}
                     \sigma & D \\  \overline D & \sigma
                \end{array}   
            \right )
\label{other-M}
\end{equation}
with $D$ and $\overline D$ given by Eq. \ref{scl-sc-D} if they are
translated to the 8-component language.  If instead of
Eq. \ref{qp-mass-mat}, the mass matrix Eq. \ref{other-M} is used, then
the zero temperature and grand potential becomes
\begin{equation}
   \Omega'/V = 4 i \int \frac{d^4p}{(2\pi)^4} \ln \left [
\left ( 1-\frac{\sigma^2} {p^2} \right ) 
\left (1-\frac{\sigma^2+\chi^2} {p^2} \right )^2 
\right ] + \frac 1 {4G_0} \sigma^2 + \frac 1 {2G_3} \chi^2.
\end{equation}
It has only one pair of minima located in the two dimensional
$(\sigma, \chi)$ plane. It has no metastable state for the vacuum
state.  This result, as it is shown in section \ref{sec:8CompTheory},
violates the mirror symmetry Eq. \ref{realcond2} as a
result of the reality condition Eq. \ref{realcond} since the scalar field
$\sigma$ couples to the fermion via a ``type--II'' vertex (see
Eq. \ref{scalar-v}) in Eq. \ref{other-M}.

\subsection{The particle spectra}

Further problems of using Eq. \ref{other-M} can be found if one study
the spectra correspond to mass matrices Eq. \ref{qp-mass-mat} and
Eq. \ref{other-M} in the presence of the chemical potential $\mu$ in
some details.  The spinor for the quasi-particle satisfies
\begin{equation}
\left (\rlap\slash p +\gamma^0\mu O_3 - \Sigma\right ) U(\mathbf{p})=0.
\label{Spinor-Eq}
\end{equation}
It has twelve solutions. For the scalar color superconductor, it can
be reduced to
\begin{eqnarray}
   (\rlap\slash p_+ - \sigma) u_1 + \gamma^5 {\cal A}_c\chi^c u_2 &=& 0,\\
   \gamma^5 {\cal A}^c \overline\chi_c u_1 + (\rlap\slash p_-\pm \sigma)u_2 &=&0
\label{8Dirac-eq}
\end{eqnarray}
corresponds Eq. \ref{qp-mass-mat} and Eq. \ref{other-M} respectively. Here
$u_1$ is the upper four component and $u_2$ is the lower four component
of $U$. 

Eq. \ref{8Dirac-eq} can be used to find the spinor $U(\mathbf{p})$ and the 
corresponding energy spectra. It is easy to shown that if the
quark has the same color as the non-vanishing $\overline \chi_c$, then 
there are four solutions
\begin{equation}
   \epsilon_\mathbf{p} = \pm E_\mathbf{p} \pm \mu
\end{equation}
with $E_\mathbf{p} = \sqrt{p^2+\sigma^2}$. The rest of the two quarks couples
to each other by the antisymmetric matrix ${\cal A}_c= -{\cal A}^c$ in
the color space. Their spectra depend on the mass matrix used. For
this pair of quarks, there are two degenerate sets of solutions each
of which contains four solutions. If mass matrix Eq. \ref{qp-mass-mat}
is used, it can be found that
\begin{equation}
     \epsilon_\mathbf{p} = \pm \sqrt{\left (E_\mathbf{p}\pm \mu \right )^2+\chi^2 
            \pm 2 \left (\sqrt{ E^2_\mathbf{p}\mu^2+\sigma^2\chi^2} - 
              E_\mathbf{p}\mu \right )}
\label{sca_sc_spt}
\end{equation}
with the $\pm$ inside the square root taking the same value for each
solution.  The four excitation modes with color different from the
order parameter $\overline\chi_c$ for mass matrix Eq. \ref{other-M}
are found to be
\begin{equation}
  \epsilon_\mathbf{p} = \pm \sqrt{\left (E_\mathbf{p}\pm \mu \right )^2+\chi^2}. 
\label{scl-spt2}
\end{equation}
It is different from Eq. \ref{sca_sc_spt} when $\mu$ is small. These two
spectra tends to each other at non-relativistic and high density limit, 
namely, $\mu>>\chi^c$ and $\sigma >> \chi^c$.

The problem with using mass matrix Eq. \ref{other-M} can be further
revealed in the $\mu\to 0$ limit in which it can be shown that the
spinor $u_1$ or $u_2$ corresponding to spectra Eq. \ref{scl-spt2} are
not constrained in any way by the Eq. \ref{Spinor-Eq}. One of them can
be arbitrarily chosen. This is not acceptable on the physical basis.

\subsection{Is $CPT$ invariance violated in the color superconducting phase?}

  According to the local theory \cite{ThPap,ThPap2}, the time
component of the statistical gauge field has non-vanishing vacuum
expectation value in the color superconducting phase $\mu^0_{vac}\ne
0$. The coupling of fermions to this vacuum induced interaction term
for fermions is
\begin{equation}
      {\cal L_{CP}}(x) = \mu^0_{vac} V_I^0(x),
\label{cp-odd-term}
\end{equation}
where the ``type--I'' current $V_I^\mu$ is given by
Eq. \ref{vec-p-tt}. This term is $CP$ odd since $V_I^0({\bf x},t)\to
-V_I^0(-{\bf x},t)$ under the $CP$ transformation. Therefore it violates
the $CP$ invariance spontaneously.

  This vacuum term is also time reversal odd in the 8-component theory
because Eq. \ref{vec-p-tt} tells us that $V_I^0({\bf x},t)\to
-V_I^0({\bf x},-t)$ under the time reversal transformation. This is
different from the 4-component theory for fermion in which the only
possible current that can be constructed to couple to the (time
component of the) statistical gauge field is even under the time
reversal transformation. But note that the existence of a
non-vanishing $\mu_{vac}^0$ depends very much on the local theory
which is formulated in the 8-component representation of fermion
field.  It can be concluded that the presence of a finite vacuum
$\mu^0_{vac}$ in the color superconducting phase of the strong
interaction vacuum state violates the time reversal also. 

   So, the $CPT$ invariance is maintained even in the possible
metastable color superconducting phase of the strong interaction
vacuum state \cite{front}. This can also be seen directly from
Eq. \ref{eq:CPT-tra} which implies that $V_I^0$ is even under the
$CPT$ transformation.  The $CPT$ invariance can be realized in two
different ways. In the first one is neither $CP$ nor $T$ is actually
violated in a process even with a term like
Eq. \ref{cp-odd-term}. This could happen since as it is discussed in
section \ref{sec:Particle} the physical amplitudes in an reaction
contain contributions from both particles and their mirror particles
which feels different $\mu_{vac}^0$ so in the final results, they
cancel each other. The second possibility is that both $CP$ and $T$
are violated. It is expected that whether the first possibility or the
second one is realized depends on the reaction that has to be sort out
in future works.

  Take the neutral kaon system for example. Since it is very likely
that the metastable color superconducting phase at zero density is
induced by a condensation of diquarks consists of light quarks (the up
and down quarks), the statistical gauge field $\mu_{vac}^0$ only
couples on the the light quarks inside the neutral kaon. Therefore
$K^0$ and $\overline K^0$ will feel opposite values of $\mu_{vac}^0$,
which leads to a violation of $CP/T$.  Taking into account of the
contribution of the mirror kaons consists of the mirror quarks, such a
$CP/T$ violating effects could be canceled in the final results if the
CPLEAR type of experiments \cite{CPLEAR} are carried out. But the
cancellation will be incomplete in the KTeV type of experiment
\cite{KTeV} due to the final state interaction effects \cite{Kaon}.

\section{Summary}
\label{sec:Sum}

It is shown in this work that it is possible to formalize the
8-component theory for fermions into a quantum field theory, which is
better behaved from the mathematical point of view in at least two
aspects: 1) the conceptual problem associated with the corresponding
quantum mechanical state of a time reversed particle is solved by
introducing a $CPT$ mirror state for each fermionic particles; 2) the
additional infinities contained in the fermionic section of a
4-component fermionic quantum field theory at finite density (or in
the presence of an external constant time component of a vector field,
which is large enough to induce the production of
particle--antiparticle pairs) are absent in the 8-component theory. It
is shown that the $CPT$ invariance can be violated in 8-component
interaction theories by certain mixing of the ``type--I'' operators
and ``type--II'' operators at the formal level. But the ``type--II''
operators are ``bad behaved'' operators in the sense they have
divergent ground state expectation values. The thermodynamics and the
canonical quantization of the theory are studied. The Feynman rules
for perturbation expansion of elementary processes and loop correction
for interacting theories are deduced. A particle and its mirror states
are mutual images of each other under the mirror reflection
transformation ${\cal M}$ at zero density.

The mirror particles for fermions introduced here is different from
the ones given in
Refs. \cite{mirror1,mirror1a,mirror1b,mirror2,mirror3}.  The mirror
world was found to be non-interacting with each other besides gravity
\cite{mirror1a}. Here, the fermion and mirror--fermion interacts with
each other in the same way that two fermions interact. In the
electro--weak sector, the left (right) handed fermions in the mirror
world of Refs. \cite{mirror1} correspond to the right (left)
handed fermions and therefore the sigh for parity violation
observables are different in sign in these two worlds. It can be shown
that here, the parity violation observables have the same sign for
fermions and the corresponding mirror fermions at zero chemical
potential. Another difference is that only fermions have mirror
partners while there is a mirror world for any particles in other
approaches, etc..  It is demonstrated that such a strongly entangled
theory for possible mirror world for any fermionic systems is possible
both theoretically and phenomenologically.

It is shown that the 8-component theory and the 4-component theory can
be made to identical at vanishing chemical potential or the time
component of a vector field when one makes a two to one mapping of the
states in the 8-component theory to a corresponding state in the
4-component theory.  Although some preliminary discussions are given
here and in an earlier work \cite{early}, the construction of and the
physics behind such a mapping at finite chemical potential or finite
time component of certain vector field remains to be explored in the
future.

It should be mentioned that the present 8-component theory can also be
applied to situations in which the time component of a background
gauge field, like the electromagnetic, gluonic or the statistical
gauge field \cite{ThPap}, is non-vanishing. These kinds of situations
arise in strong interaction at the mean field level, in the vicinity
of a small charge carrier like a heavy ion \cite{Grein}, or in
numerical simulations of gauge field where all kinds of relevant gauge
field configurations are generated.

\section*{Acknowledge}

This work is supported by the National Natural Science Foundation of China
under the contract 19875009. I would like to thank Dimiter G. Chakalov for
communications.







\end{document}